\documentclass[nofootinbib,aps,prd,reprint,superscriptaddress,showkeys]{revtex4-2}

\usepackage[utf8]{inputenc} 
\usepackage{graphicx,overpic,mathtools}
\usepackage{amsthm,amsmath,amssymb,hyperref}
\usepackage{braket,bm,bbm,setspace}
\usepackage[normalem]{ulem} 
\usepackage{physics}
\usepackage{float}
\usepackage[makeroom]{cancel}
\usepackage[english]{babel}
\usepackage{xcolor}
\usepackage{tensor}
\graphicspath{ {images/} }
\addto\captionsspanish{}
\hypersetup{
colorlinks=true,
pdfborder={0 0 0},
citecolor=purple,
linkcolor=blue,
filecolor=blue,
urlcolor=blue,
}
\usepackage{subfigure}

\usepackage{enumerate}
\usepackage{orcidlink}

\usepackage{multirow}

\usepackage{tikz}
\usetikzlibrary{quantikz2}

\makeatletter
\def\th@plain{
\thm@notefont{}
\itshape
}
\def\th@definition{
\thm@notefont{}
\normalfont
}
\makeatother

\begin{document}

\newcommand{\ci}{\text{i}}
\newcommand{\identity}{\mathbbm{1}}
\newcommand{\TM}{G}

\newcommand{\update}{V}

\newcommand{\apriori}{\textit{a priori} }
\newcommand{\Apriori}{\textit{A priori} }

\newcommand{\qpestate}{\xi}

\title{Efficient Simulation of Szegedy Quantum Walk Formulations and Algorithms}
\author{Sergio A. Ortega \orcidlink{0000-0002-8237-7711}}
\email{s.a.ortega@yonsei.ac.kr}
\affiliation{Department of Statistics and Data Science, Yonsei University, Seoul 03722, Republic of Korea}
\author{Daniel K. Park \orcidlink{0000-0002-3177-4143}}
\email{dkd.park@yonsei.ac.kr}
\affiliation{Department of Statistics and Data Science, Yonsei University, Seoul 03722, Republic of Korea}
\affiliation{Department of Applied Statistics, Yonsei University, Seoul 03722, Republic of Korea}
\affiliation{Department of Quantum Information, Yonsei University, Seoul 03722, Republic of Korea}

\begin{abstract}
	{Quantum walks provide a versatile framework for quantum algorithms across a wide range of applications. We develop efficient classical simulation methods for Szegedy quantum walks that avoid explicit construction of the full unitary evolution operator. Unlike previous approaches restricted to a particular walk formulation, our framework is built from fundamental update and reflection operators, enabling the simulation of a broader class of Szegedy walk formulations. We further extend these methods to phase-estimation-based algorithms coupled to the walk, including implementations suitable for large sparse graphs. The resulting methods achieve optimal $O(N^2)$ complexity for dense graphs with $N$ nodes. For sparse graphs, the computational cost scales linearly with the number of edges, which is $O(N)$ in many cases. We implement the framework in the Python package SQWLib and illustrate its capabilities through simulations of representative algorithms, including quantum simulated annealing and quantum search on graphs. These results provide a practical tool for studying Szegedy-walk-based algorithms numerically beyond purely analytical treatments.}
\end{abstract}


\maketitle

\onecolumngrid
\section{Introduction}\label{Introduction}

Quantum walks constitute a fundamental framework in quantum computing, providing algorithmic capabilities beyond those of classical random walks across a wide range of computational tasks. They were first introduced in discrete time \cite{QRW}, and later in continuous time \cite{Trees}. An important discrete-time quantum walk is the one introduced by Szegedy \cite{Szegedy}, which is a generalization of the coined Grover walk \cite{Sandbichler-master,Graph-phased}. This approach is useful for quantizing arbitrary Markov chains.

As a naive quantum walk, i.e., the unitary evolution of a quantum state representing the position of a walker on a graph, it has been shown to be useful for problems of optimization \cite{Lemieux,Qfold,QMS,GWQMA,Linear}, classification \cite{Paparo1,Paparo2,APR}, quantum search \cite{Portugal,Searchrank,S_queries}, and blockchain technologies \cite{Bitcoin}. However, this quantum walk can also be coupled with quantum phase estimation (QPE) to provide algorithms closely related to machine learning. One such QPE-based algorithm is quantum simulated annealing (QSA) \cite{Boixo}, which is used to obtain samples of the classical stationary distribution of Markov chains with speedup. This algorithm can thus be used for Boltzmann sampling, and could accelerate machine learning algorithms such as Boltzmann machines \cite{BM,Gilhan}. Another interesting algorithm is quantum search on graphs via approximate amplitude amplification \cite{Santha}, which has been proposed as a subroutine for speeding up the decision-making process of classical reinforcement learning agents \cite{Paparo3}.

Despite their theoretical relevance, the practical study of Szegedy quantum walks remains limited to theoretical analyses of specific problems. Thus, there is a lack of simulation results validating such findings. Moreover, numerical simulation is a fundamental tool for developing heuristics quantum algorithms whose complexity cannot be studied analytically. Recently, a simulation framework for the Szegedy quantum walk that avoids the explicit construction of the unitary evolution matrix was developed, giving rise to the Python package SQUWALS \cite{Squwals}. This algorithm was shown to be efficient and optimal, scaling as $\mathcal{O}(N^2)$ for dense graphs with $N$ nodes, thus improving upon previous simulation methods scaling as $\mathcal{O}(N^3)$. However, this framework can only simulate the naive quantum walk unitary evolution on a given state, but not algorithms based on QPE. Moreover, although additional formulations have appeared since Szegedy's original work, this simulation framework is only suitable for the original version \cite{Szegedy,Notes}.

In this work, we address this gap by developing a more general classical simulation framework for Szegedy quantum walks and related algorithms. Our main contributions are as follows:

\begin{enumerate}[i)]
    \item We introduce efficient simulation methods for fundamental operators underlying Szegedy quantum walks, in particular the update and reflection operators. This operator-level formulation enables the simulation of a broader class of Szegedy walk constructions beyond the standard formulation, while retaining optimal quadratic scaling for dense graphs.

    \item We extend this framework to sparse graphs through efficient sparse-state representations, reducing the computational cost to linear scaling in the number of edges and enabling simulations on large sparse instances.

    \item We develop an explicit and implementable simulation framework for QPE-based Szegedy-walk algorithms without constructing the full unitary matrix, thereby making it possible to study algorithms that go beyond direct walk evolution.

    \item We implement these methods in the Python package SQWLib and demonstrate their applicability through representative simulations, including marked-node detection, quantum simulated annealing, and quantum search on graphs.
\end{enumerate}

The paper is organized as follows. In Section \ref{sec:Szegedy}, we review the formulation of the Szegedy quantum walk. In Section \ref{sec:Simulation}, we present efficient classical algorithms for simulating this quantum walk. In Section \ref{sec:QPE}, we review the quantum phase estimation algorithm and show how it can be simulated within our framework. In Section \ref{sec:Results}, we present SQWLib, our Python simulator for the quantum walk and QPE, and provide illustrative results. Finally, in Section \ref{sec:Conclusions}, we summarize our findings and conclude.

\section{Szegedy Quantum Walk}\label{sec:Szegedy}

\subsection{Original Formulation}\label{sec:Original}

A classical Markov chain can be represented as a random walk occurring on the nodes of a graph. For a graph with $N$ nodes, the chain has an associated $N \times N$ transition matrix $G$, whose elements $G_{ji}$ are the probabilities of the walker jumping from node $i$ to node $j$. The Szegedy quantum walk provides a quantization of this Markov chain \cite{Szegedy}. The Hilbert space of the quantum system is composed of two $N$-dimensional registers, where the first one usually represents the position of the walker, and the second one corresponds to a inner degree of freedom indicating the possible directions a walker can take. This inner state is usually referred as coin state, since the directions are determined by the quantum equivalent of tossing a multidimensional coin in a random walk. Thus, it actually corresponds to the span of all the vectors representing the $N \times N$ directed edges:
\begin{equation}\label{Hilbert}
	\mathcal{H} = \text{span}\lbrace\left|i\right>_1\left|j\right>_2,\ i,j = 0,1,...,N-1\rbrace = \mathbb{C}^N \otimes \mathbb{C}^N.
\end{equation}
In this paper we count the nodes of the network, and therefore the matrix indexes, from $0$ to $N-1$.

The classical transition matrix is used to associate each node with a state $\left|\psi_i\right>$ representing a superposition of all the edges outgoing from the $i^{th}$ vertex:
\begin{equation}\label{psi_i}
	\left|\psi_i\right> := \left|i\right>_1 \otimes \left|\omega_i\right>_2, \ \ \ \ \left|\omega_i\right> = \sum_{k=0}^{N-1} \sqrt{\TM_{ki}}\left|k\right>.
\end{equation}
Similar to a coined quantum walk, the Szegedy quantum walk makes uses of a coin-like operator that determines the possible directions of the walker, and a shift that performs the movement \cite{Notes}. For the coin, we define a reflection $R$ around the space spanned by the $\left|\psi_i\right>$ states:
\begin{equation}\label{reflection}
	R = 2\Pi - \mathbbm{1},
\end{equation}
where the projector is defined as:
\begin{equation}\label{projector}
\Pi := \sum_{i=0}^{N-1} \left|\psi_i\right>\left<\psi_i\right|.
\end{equation}
For the shift, we make use of a swap operator that exchanges the information between both registers:
\begin{equation}\label{swap}
	S := \sum_{i,j=0}^{N-1} \left|i,j\right>\left<j,i\right|.
\end{equation}
From now on, we use the notational convention that the tensor product of two register states may be written as a single ket with two indices separated by a comma; e.g., $|i,j\rangle \equiv |i\rangle\otimes |j\rangle$. Combining the reflection $R$ and the swap $S$, we obtain the unitary evolution operator of the walk:
\begin{equation}\label{U}
	U := SR.
\end{equation}

Although this formulation is commonly used in the recent literature \cite{Paparo1,Paparo2,Semiclassical,Squwals}, the first formulation of the evolution operator proposed by Szegedy \cite{Szegedy} was a product between two reflections, $W=R_BR_A$, where $R_A$ corresponds to the reflection around the states $\left|\psi_i\right>$, and $R_B$ around the swapped subspace. Thus, $R_A = R$ and $R_B = SRS$, so that:
\begin{equation}\label{W}
	W = R_BR_A = SRSR = U^2.
\end{equation}
For this reason, we refer to $U$ as the single-step Szegedy operator, and $W$ as the double-step Szegedy operator.

The initial state of the system is chosen as a linear combination of the $\left|\psi_i\right>$ states, whose coefficients are given by the equivalent initial classical probability distribution. Doing so, the evolution of the quantum walk is restricted to the following subspace \cite{Paparo1}:
\begin{equation}\label{Dynamical_subspace}
	\mathcal{H}_D := \text{span}\left\lbrace\left|\psi_i\right>,S\left|\psi_i\right>,\ i = 0,1,...,N-1\right\rbrace,
\end{equation}
This subspace is usually referred as dynamical subspace. This nomenclature comes from the fact that for the operator $W$, all the dynamical eigenvalues, i.e., those different to $1$, are contained in this subspace. Thus, for an arbitrary state, the perpendicular component outside this subspace remains static, and only the parallel component has dynamics. An important feature for reversible Markov chains, as those obtained from the Metropolis-Hastings algorithm \cite{Metropolis,Hastings}, is that this subspace has a unique eigenvalue $1$, whose eigenstate is
\begin{equation}\label{pi}
	\left|\pi\right> := \sum_{k=0}^{N-1} \sqrt{\pi_i}\left|\psi_i\right> = \sum_{k=0}^{N-1} \sqrt{\pi_i}\left|i\right>_1 \otimes  \left|\omega_i\right>_2,
\end{equation}
where $\pi$ is the stationary distribution of the classical Markov chain. Therefore, a measurement in the first register provides a sample of the classical stationary distribution. This fact is important for QPE-based machine learning algorithms that aim to sample the stationary distribution, as the ones we explore in Section \ref{sec:Results}.

Finally, although in many quantum walk algorithms the probability distribution is obtained by measuring the first register, it is worth mentioning that there are also algorithms where the information of interest is obtained by measuring the second register instead, as for example the quantum PageRank and SearchRank \cite{Paparo1,Paparo2,APR,Searchrank,Randomized}.

\subsection{Other Szegedy Walk Formulations}\label{sec:Alternative}

Apart from the previous formulation of the walk, it has emerged different formulations that preserve the spectral properties, but provide different features suitable for particular algorithms, as for examples those aimed to quantum simulated annealing for optimization \cite{Boixo,KMOR15,Lemieux}. In order to obtain these alternative formulations, first let us introduce the update operator $\update$. This is defined by its actions as:
\begin{equation}\label{update}
	\update\left|i\right>_1\left|0\right>_2 = \left|i\right>_1\left|\omega_i\right>_2 = \left|\psi_i\right>,
\end{equation}
so that it creates the $\left|\psi_i\right>$ states reading the computational basis state of the first register, provided the second register is in the computational basis state $\left|0\right>$. Let us also define a reflection around the states $\left|i\right>_1\left|0\right>_2$ as:
\begin{equation}
	R_0 = 2\sum_{i=0}^{N-1} \left|i,0\right>\left<i,0\right| - \identity.
\end{equation}
This reflection and the update operators can be used as more fundamental operators, since their composition allow to obtain the previous reflection $R$ in \eqref{reflection} as $R = \update R_0 \update^\dagger$. Note that the update operator is not uniquely defined, since we only provide the action on $N$ states of the total $N^2$-dimensional space. However, since whichever definition provides the same reflection $R$ since the non-definition effect is canceled when $V$ and $V^\dagger$ encounters around $R_0$, this makes the action on the rest of states \apriori irrelevant, at least, for the original formulation of the walk. We examine in Section \ref{sec:Comparison} how this actually affects the results of different formulations.

From these more fundamental operators, we can then express the original Szegedy operator $W = SRSR$ as:
\begin{equation}
	 W = S \update R_0 \update^\dagger S \update R_0 \update^\dagger.
\end{equation}
With this operator, a similarity transformation yields an alternative evolution operator \cite{Boixo,Boixo_2}:
\begin{equation}\label{W_2}
	\widetilde{W} = \update^\dagger W \update = \update^\dagger S \update R_0 \update^\dagger S \update R_0.
\end{equation}
Note that this operator actually corresponds to a rearrangement of the original one, where the first inverse update operator $V^\dagger $ is moved to the end.

Since now the walk has a different structure, the initial state has also to be redefined accordingly. The first operator that acts on a state is the reflection $R_0$ around the states $\left|i,0\right>$, rather than around the states $\left|\psi_i\right>$. Thus, the initial state must be chosen as a linear combination of these $\left|i,0\right>$ states. Moreover, given the relationship $\left|i,0\right>= \update^\dagger\left|\psi_i\right>$, note that the walk would perform the same evolution as the original formulation, with an additional application of the inverse of the update operator $\update^\dagger$ at the end. However, note that in this case the last inverse update operator $\update^\dagger$ does not find its partner to cancel the non-definition effect, so that unless the state of the walk before the last $\update^\dagger$ lies in the subspace spanned by the $\left|\psi_i\right>$ states, the action of $\update^\dagger$ is not uniquely defined and depends on the particular implementation. A natural choice for a quantum circuit implementing $V$ would always be constructed by gates acting in the second register controlled by the first one. Therefore, the information of the first register is only read to modify the state on the second one, but is unaltered. Thus, the probability distribution sampled from the first register, which usually corresponds to the position of the walker, remains the same, despite the quantum amplitudes of the final state being implementation dependent.

Although we have defined this alternative formulation of the walk directly from the similarity transformation, it is usually defined in an analog manner as the product of two reflections, so that $\widetilde{W} = \widetilde{R}_A\widetilde{R}_B$, with
\begin{equation}
	\widetilde{R}_A = R_0,
\end{equation}
\begin{equation}
	\widetilde{R}_B = \update^\dagger S \update R_0 \update^\dagger S \update.
\end{equation}
With this interpretation, it is possible to use the theorem of Szegedy for a product of two reflections \cite{Szegedy,Boixo,Boixo_2}, to obtain that in this case in the dynamical subspace where the walk takes place is
\begin{equation}
	\mathcal{H}_{\widetilde{D}} := \text{span}\left\lbrace\left|i,0\right>,\update^\dagger S \update\left|i,0\right>,\ i = 0,1,...,N-1\right\rbrace.
\end{equation}
Again, this subspace depends on the transition matrix, and moreover, it is not uniquely defined due to the action of the inverse of the update operator $\update^\dagger$ after the swap $S$. However, this is \apriori irrelevant for an algorithm based on the classical stationary distribution of reversible Markov chains, since in this case again there is only a single eigenvalue $1$ associated to it by the transition-matrix-independent state
\begin{equation}\label{pi_2}
	\left|\widetilde{\pi}\right> := \sum_{k=0}^{N-1} \sqrt{\pi_i}\left|i\right>_1 \left|0\right>_2.
\end{equation}
The fact that this state is independent of the transition matrix is useful for quantum walks whose transition matrix changes along the steps. For example in an annealing process the stationary distribution is obtained at each step for a different temperature, and since this state is independent of the transition matrix, Markov chains coming from different temperature values can be applied sequentially (for further details see the QSA algorithm in Section \ref{sec:QSA}).

Apart from this particular construction that we have shown, the formulation of the walk in terms of the update operator $\update$ allows further constructions by rearranging the fundamental operators of the walk. For example, in the context of constructing efficient circuits for quantum Metropolis-Hastings algorithms and studying heuristic annealing algorithms \cite{Lemieux}, the following operator of the walk is studied:
\begin{equation}
	U = R_0 \update^\dagger S \update.
\end{equation}

\section{Quantum Walk Simulation of Operators}\label{sec:Simulation}

The Hilbert space of the Szegedy quantum walk $\mathcal{H}$ is of dimension $N^2$. Thus, in order to simulate the walk with a classical computer it would be in principle required to construct a $N^2 \times N^2$ matrix representing the unitary evolution operator. This operator has actually $\mathcal{O}(N^3)$ non-null elements for arbitrary transition matrices, so that using a sparse representation of the matrix would provide a classical simulation algorithm scaling as $\mathcal{O}(N^3)$ in both time and memory. Recently, an algorithm that avoids the explicit construction of this matrix was proposed, with an optimal scaling of $\mathcal{O}(N^2)$ for dense transition matrices, giving rise to the Python package SQUWALS \cite{Squwals}. This simulator provides methods for simulating the reflection $R$ and the swap $S$ operators, as well as oracles similar to the one of the Grover algorithm \cite{Grover,S_queries} applied to any register. Thus, it is suitable for the original formulation of the walk explained in Section \ref{sec:Original}. However, as we have seen in the previous section, the battery of quantum-walk-based algorithms can be extended with a reformulation in terms of more fundamental operators. Therefore, we need methods for simulating also the update operator $\update$ and the reflection $R_0$ if we are to study these further algorithms. With this aim, in this section we extend the methods in Ref. \cite{Squwals} for simulating these fundamental operators.

\subsection{Memory-Saving Algorithms}

To simulate the quantum walk with the minimum requirements possible, the key is simulating the action of the operators avoiding their explicit matricial construction. To do so, note that an arbitrary quantum state in the Hilbert space $\mathcal{H}$ can be expressed as:
\begin{equation}\label{vector}
	\left|\phi\right> = \sum_{i,j=0}^{N-1} a_{ij} \left|i\right>_1 \left|j\right>_2.
\end{equation}
Since the amplitudes correspond to tensor products of two computational basis states, we can represent them as a $N\times N$ matrix $\Phi$ rather than a vector, whose elements are
\begin{equation}\label{PHI}
	\Phi_{ij} = a_{ji}.
\end{equation}
For convenience, the column index represents the first register, whereas the row index represents the second register. Thus, if we divide the vector state into blocks corresponding to each state of the computational basis of the first register, then each block corresponds to each column of the matrix state.

Given the matrix representation of a state, it turns out that the action of the swap operator $S$ corresponds to transposing this matrix. Regarding the reflection $R_0$ around the states $\left|i,0\right>$, since this operator does not depend on the transition matrix, its implementation is also straightforward. We just have to multiply by $-1$ all rows of the state matrix except the first one. However, the same as happens in Ref. \cite{Squwals}, the difficult part comes from the operator that depends on the transition matrix $\TM$, which in this case is the update operator $\update$. From a circuital perspective, any quantum circuit construction satisfying \eqref{update} would be enough, since it would be unitary and therefore correct. However, for general graphs, constructing such circuits is not easy. Moreover, in this work we are interested in numerical classical simulations, so we need to explicitly give a particular construction of the operator.

In the following sections, we show two different methods to simulate particular implementations of the update operator $\update$ and its inverse $\update^\dagger$, which not being unitary equivalent, also allow to check whether this fact is actually irrelevant or not for due algorithms.

\subsubsection{Update operator 1}

A form of simulating the update operator $\update$ is to look for the minimal subspace where the action of the operator is not trivial, and define it as the identity for the orthogonal complement. As we will see, this subspace is not as trivial as a span of computational basis states.

Given the mapping in equation \eqref{update}, we define the subspace where the action is not trivial as:
\begin{equation}
	\mathcal{H}_\update := \text{span}\lbrace{\left|i\right>_1\left|0\right>_2,\left|\psi_i\right>,\ i = 0,1,...,N-1\rbrace},
\end{equation}
and denote it as the update subspace. The action on $\mathcal{H}^\perp_\update$ must be then the identity. Moreover, since neither the update operator nor its inverse affect the first register, this subspace can be split into $N$ invariant subspaces, each of them indexed by index $i$.\\

Note that the states generating the subspace $\mathcal{H}_\update$ in general do not form a orthonormal basis. For example, in the case of nodes that only point to node $0$, so that $\left|\psi_i\right> = \left|i\right>_1\left|0\right>_2$, the generator set has redundant vectors. In this case, the corresponding subspace is trivial and for that index $i$ the update operator acts as the identity on any computational basis state with $i$ in the first register. Therefore, to develop a mathematical expression of the operator, let us for the moment consider that all these subspaces are 2-dimensional, and thus $\mathcal{H}_\update$ is of dimension $2N$.

In the simple case that a node does not point to node $0$, then its two corresponding generator states are perpendicular, and directly form a orthonormal basis of the corresponding subspace. However, let us consider the general case where they are not perpendicular. For each of the $N$ subspaces, we have two forms of constructing an orthonormal basis, taking one of the vectors and expanding with a perpendicular one. Let us first take the vector $\left|i\right>_1\left|0\right>_2$, and define a perpendicular vector in the subspace as the normalized support of the state $\left|\psi_i\right>$ on the rest of computational basis states:
\begin{equation}
	\left|i\right>_1\left|0^\perp\right>_2 = \frac{\displaystyle \left|i\right> \otimes \sum_{k=1}^{N-1} \sqrt{\TM_{ki}}\left|k\right>_2}{\displaystyle \sqrt{\sum_{k=1}^{N-1} \TM_{ki}}},
\end{equation}
so that we obtain the basis
\begin{equation}
	\mathcal{B}_1 = \lbrace{\left|i\right>_1\left|0\right>_2,\left|i\right>_1\left|0^\perp\right>_2,\ i = 0,1,...,N-1\rbrace}.
\end{equation}
The other form is taking the states $\left|\psi_i\right>$ and expanding with perpendicular states. To do so, we expand it in the basis $\mathcal{B}_1$, so that we have a two-dimensional vector, and we obtain the corresponding perpendicular partner swapping the components and changing the sign of one of them:
\begin{equation}\label{psi_i_perp}
	\left|\psi_i^\perp\right> = \left<i,0^\perp\right|\left.\psi_i\right>\left|i,0\right> - \left<i,0\right|\left.\psi_i\right>\left|i,0^\perp\right> = \sqrt{\sum_{k=1}^{N-1} \TM_{ki}}\left|i,0\right> - \TM_{0i}\left|i,0^\perp\right>,
\end{equation}
so that
\begin{equation}
	\mathcal{B}_2 = \lbrace{\left|\psi_i\right>,\left|\psi_i^\perp\right>,\ i = 0,1,...,N-1\rbrace}.
\end{equation}

Now, we can define the unitary of the update operator so that it goes from the states of $\mathcal{B}_1$ to the states of $\mathcal{B}_2$, and is the identity for any state in the orthogonal complement. A simulation strategy for the action on a general state would be to project this state into the update subspace $\mathcal{H}_\update$, so that we obtain a parallel and a perpendicular component. We expand the parallel component in the basis $\mathcal{B}_1$, and recompose the state with the same coefficients applied to the basis $\mathcal{B}_2$ to transform it, and finally we add the unaltered perpendicular component. Since the perpendicular component is the subtraction between the original state and its parallel component, the update operator can be expressed as:
\begin{equation}
	\update = \sum_{i=0}^{N-1} \left|\psi_i\right>\left<i,0\right| + \sum_{i=0}^{N-1} \left|\psi_i^\perp\right>\left<i,0^\perp\right| + \identity - \sum_{i=0}^{N-1} \left|i,0\right>\left<i,0\right| - \sum_{i=0}^{N-1} \left|i,0^\perp\right>\left<i,0^\perp\right|.
\end{equation}
It is easy to check that this operator acts as desired. For any perpendicular state only the identity acts. However, for any state in $\mathcal{H}_\update$, the last two terms, corresponding to the projector onto $\mathcal{H}_\update$, let the state unchanged and thus are canceled out with the identity, so that only the first two terms, which produce the transition between basis, act. For the inverse, we analogously obtain:
\begin{equation}
	\update^\dagger = \sum_{i=0}^{N-1} \left|i,0\right>\left<\psi_i\right| + \sum_{i=0}^{N-1} \left|i,0^\perp\right>\left<\psi_i^\perp\right| + \identity - \sum_{i=0}^{N-1} \left|\psi_i\right>\left<\psi_i\right| - \sum_{i=0}^{N-1} \left|\psi_i^\perp\right>\left<\psi_i^\perp\right|.
\end{equation}
Note that the last two terms, when added, are equal to the added last two terms of $\update$. We express them this way for simulation convenience.

For simulating numerically this operator, we want to use only element-wise operations from the transition matrix, following a similar procedure to the simulation of the reflection operator $R$ in the library SQUWALS \cite{Squwals}. The strategy for the update operator is to calculate the scalar products with the states of the basis $\mathcal{B}_1$, and use them to reconstruct the action of each of the terms. Therefore, we first need to represent these basis states.

\begin{figure}[t]
	\centering
	\includegraphics[scale=0.46]{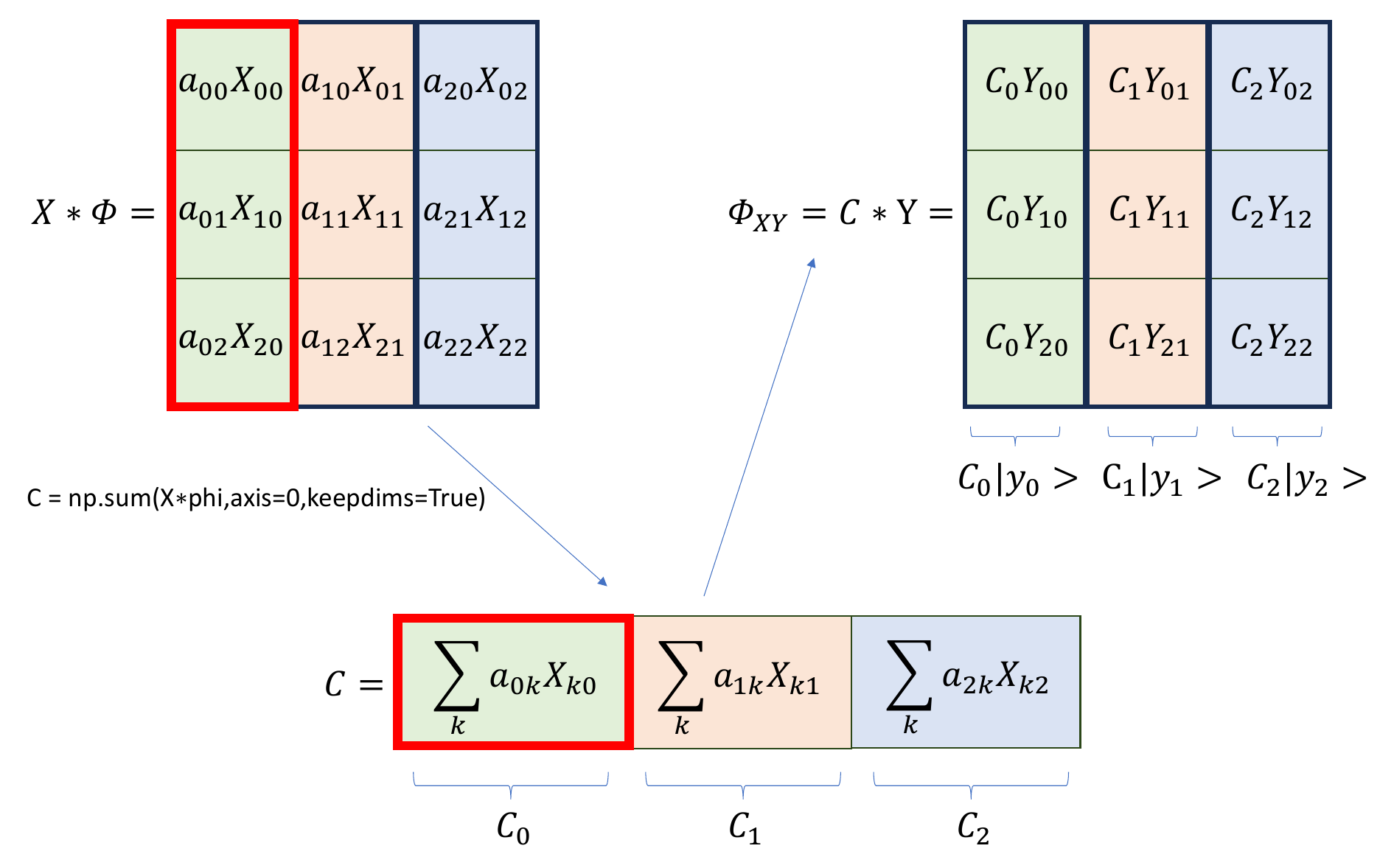}
	\caption{Expansion-recomposition example for a graph with $N=3$ nodes. For the expansion in the set of states $\left|x_i\right>$, the coefficients are calculated by the element-wise multiplication of the matrix $X$ and the matrix state $\Phi$, and then adding over the rows. For the recomposition on a set of states $\left|y_i\right>$, the coefficients vector $C$ is multiplied element-wise with each row of the matrix $Y$.}
	\label{F:expansion-recomposition}
\end{figure}

Before proceeding, let us note that apart from the trivial identity term, all the terms have the following structure:
\begin{equation}
	\Pi_{XY} = \sum_{i=0}^{N-1} \left|y_i\right>\left<x_i\right|,
\end{equation}
so that they are operations composed of the following two primitives:
\begin{itemize}
	\item Expansion: It consists of obtaining the coefficients of a state $\left|\phi\right>$ along a set of states $\left|x_i\right>$:
	\begin{equation}
		C_i = \left<x_i\right.\left|\phi\right>.
	\end{equation}
	\item Recomposition: It consists of using some coefficients for reconstructing a state given a set of states $\left|y_i\right>$:
	\begin{equation}
		\Pi_{XY}\left|\phi\right> = \left|\phi\right>_{XY} = \sum_{i=0}^{N-1} C_i\left|y_i\right>.
	\end{equation}
\end{itemize}
Moreover, all these states have a similar structure:
\begin{equation}
	\left|x_i\right> := \left|i\right>_1 \otimes \sum_{k=0}^{N-1} X_{ki}\left|k\right>_2.
\end{equation}
For the Reflection operator $R$ in \eqref{reflection} we would have $\left|x_i\right> = \left|y_i\right> = \left|\psi_i\right>$, in the projector operator $\Pi$. In our case we can follow a similar simulation procedure of that of Ref. \cite{Squwals} for this general term. Since the states $\left|x_i\right>$ have in the first register only the computational basis states, they have a block form where only the block indexed by $i$ has non-null elements. Therefore, we can just store all the non-null elements of the set in a $N\times N$ matrix denoted as $X$, where each column corresponds to each state $\left|x_i\right>$. The same applies for the set $\left|y_i\right>$. From this matrix, we can perform the expansion primitive, i.e, all the scalar products at the same time with a general state, just multiplying the matrices element-wise, and adding over the rows, as shown in Figure \ref{F:expansion-recomposition}. Once obtained the coefficients, we use the broadcasting feature of NumPy \cite{NumPy} to multiply this vector by the matrix $Y$. The result corresponds to the matrix representing the recomposed state as a linear combination of the $\left|y_i\right>$ states, as also shown in Figure \ref{F:expansion-recomposition}.

Now, we need to obtain the matrices representing the elements of both basis $\mathcal{B}_1$ and $\mathcal{B}_2$, which intervene in both operators. However, note that for the states $\left|i,0\right>$ we actually do not need it. The expansion is trivially done just taking the first row of the the matrix state $\Phi$, and the recomposition on these states is done putting due coefficients in the first row of the recomposed state. Thus, we only need the matrices for the states $\left|\psi_i\right>$, $\left|\psi_i^\perp\right>$ and $\left|i,0^\perp\right>$.

The first step is to obtain the matrix $\Psi$ representing all the $\left|\psi_i\right>$ states, which is obtained taking the element-wise square root of the transition matrix $\TM$ \cite{Squwals}. From this matrix, we can create the matrix of the perpendicular states $\left|i,0^\perp\right>$ in $\mathcal{B}_1$ just putting at $0$ the first row, and normalizing each column at norm 2. To do so, we take the element-wise square modulus of each element, add over the rows, and divide each column by the square root of the results. Finally, the matrix of the states $\left|\psi_i^\perp\right>$ can also be obtained with the previous primitives by a expansion-recomposition of the states $\left|\psi_i\right>$ on the basis $\mathcal{B}_1$ following equation \eqref{psi_i_perp}. Indeed, the coefficients for the recomposition with $\left|i,0\right>$ are trivially given by the first row of $\Psi$, and for the recomposition with $\left|i,0^\perp\right>$ they are the norms obtained before when normalizing the states $\left|i,0^\perp\right>$.

After having obtained these three matrices, we can use them to simulate the action of each term of the operators with the expansion-recomposition procedures already explained.

\subsubsection{Update operator 2}

Whereas the previous formulation is in some sense quite clean, since it is just the identity on a $(N^2-2N)$-dimensional subspace, its implementation requires two expansion and four recomposition procedures. Another more efficient implementation of the update operator comes from the construction of Householder reflections \cite{Isometry}. Since the action defining the update operator is an isometry, we can define a vector for each of the $N$ coin states $\left|\omega_i\right>$ of the second register as:
\begin{equation}
	\left|v_i\right> = \frac{\left|\omega_i\right>-\left|0\right>}{\left|\left|\left|\omega_i\right>-\left|0\right>\right|\right|},
\end{equation}
and use them to construct unitaries that create each of the coin states from the state $\left|0\right>$ as:
\begin{equation}
	\update_i = \identity - 2\left|v_i\right>\left<v_i\right|.
\end{equation}
The update operator is constructed as a block-diagonal matrix where each block is $\update_i$, and thus, it can be expressed as:
\begin{equation}
	\update = \identity - 2\sum_{i=0}^{N-1}\left|i,v_i\right>\left<i,v_i\right|.
\end{equation}
The same as before, we have states with support only on the block corresponding to each of the computational basis states of the first register, so that we can represent all of them at the same time with a single matrix. Indeed, this matrix is obtained by subtracting $1$ to the first row of the matrix $\Psi$, and then normalizing the columns at norm $2$. To simulate the projector in the second term of $\update$, we can use again the expansion-recomposition procedure shown in Figure \ref{F:expansion-recomposition}, and after that we can simply simulate the action of $\update$. Regarding its inverse, note that this operator is self-inverse by construction, so that there is no need to separately implement a method for $\update^\dagger$.

\subsection{Dense and Sparse Computational Complexity}

In the simulation methods provided above, the biggest object that intervene are $N \times N$ matrices, since we avoid the construction of explicit $N^2\times N^2$ unitary operators. Moreover, our methods take advantage of element-wise operations, rather than matrix multiplications. Thus, at each operation, each element of the matrices intervene only once. For these reason, if we store these matrices as dense objects, we can state that the memory and time requirements of the classical simulation scale as $\mathcal{O}(N^2)$.

This complexity is optimal for dense graphs, since the minimal information required to represent it is the transition matrix $\TM$, which also has $N^2$ elements. Nevertheless, a lot of graphs of interest are quite sparse, so that each node connect only to a small set of nodes, and thus the transition matrix has $\mathcal{O}(N)$ non-null elements. Thus, let us analyze what are the minimal requirements of our algorithm if we only store the non-null elements of the objects.

Since the Hilbert space $\mathcal{H}$ is composed by states representing the edges of the graph, it is natural that any quantum state has no support in edges that are not present in the graph. Thus, \apriori we could think of a subspace formed by the directed  edge states $\left|i,j\right>$ with a non-null transition probability $G_{ji}$. However, due to the swap operator $S$, as long a directed edge state $\left|i,j\right>$ appears in a walk state, its swapped arc $\left|j,i\right>$ also appears. Therefore, to form this reduced subspace, we must take into account the backbone of the underlying undirected graph. Let $A$ be the adjacency of the undirected backbone, so that $A_{ij} = 1$ if nodes $i$ and $j$ are connected in any direction, i.e., either $G_{ij} > 0$ or $G_{ji} > 0$, and $A_{ij} = 0$ if both transition probabilities are null. Then, the reduced subspace where the Szegedy walk takes place is defined as \cite{Graph-phased}:
\begin{equation}\label{reduced}
	\mathcal{H}_R := \text{span}\lbrace{\left|i\right>_1\left|j\right>_2 : A_{ji} = 1\rbrace}.
\end{equation}
Therefore, the mathematical object representing an arbitrary state shares the same sparsity that the symmetrized version of the transition matrix $\TM$.

For the original walk formulation, all the objects would share this feature, so that the memory requirements, as well as the time given by the number of operations, grows linearly with the number of undirected edges of the graph. However, in the case of the update operator $\update$, recall that we have also edge states of the form $\left|i\right>_1\left|0\right>_2$. Thus, for the simulation of this operator, we must extend the undirected backbone graph with edges connecting node $0$ to all the nodes. Nevertheless, this is just an addition of at most $2N$ elements, so that the complexity $\mathcal{O}(N)$ for sparse graphs is not compromised.

Although a straightforward implementation with a package able to manage sparse matrices like SciPy \cite{Scipy} would be enough for the walk simulation on a state, in the Supplemental Material (SM) we also provide a low-level method for implementing the sparse operations directly with NumPy \cite{NumPy}. This allows the vectorization of the operations for simulating the evolution on batches of states. Moreover, all the states are aligned in the reduced subspace $\mathcal{H}_R$ thanks to significant zeros, allowing the application of transformations required for the QPE simulation. See Supplementary Material \cite{SM} for further information.

\subsection{Testing the Implementation Dependence}\label{sec:Comparison}

\begin{figure*}[h]
	\centering
	\subfigure[]{\includegraphics[scale=0.5]{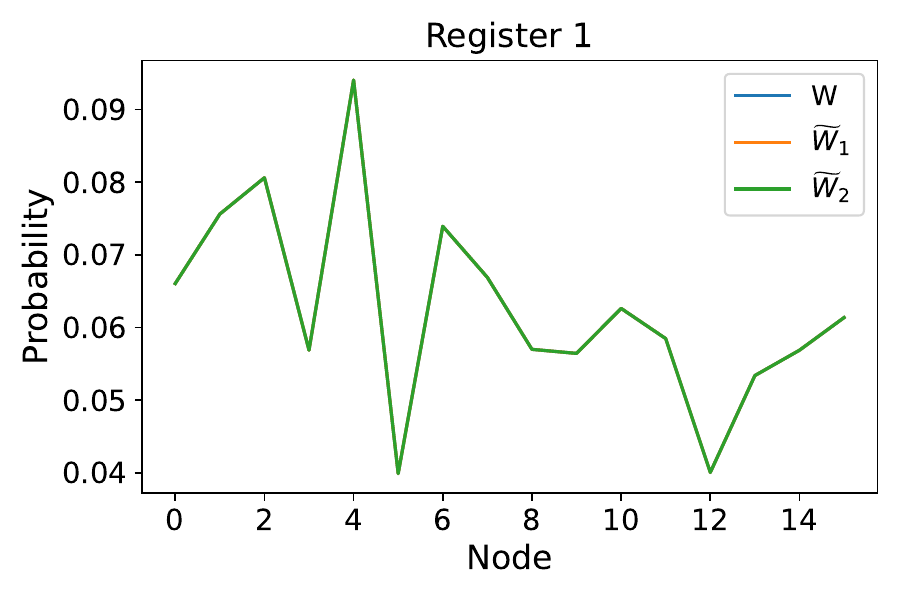}}
	\subfigure[]{\includegraphics[scale=0.5]{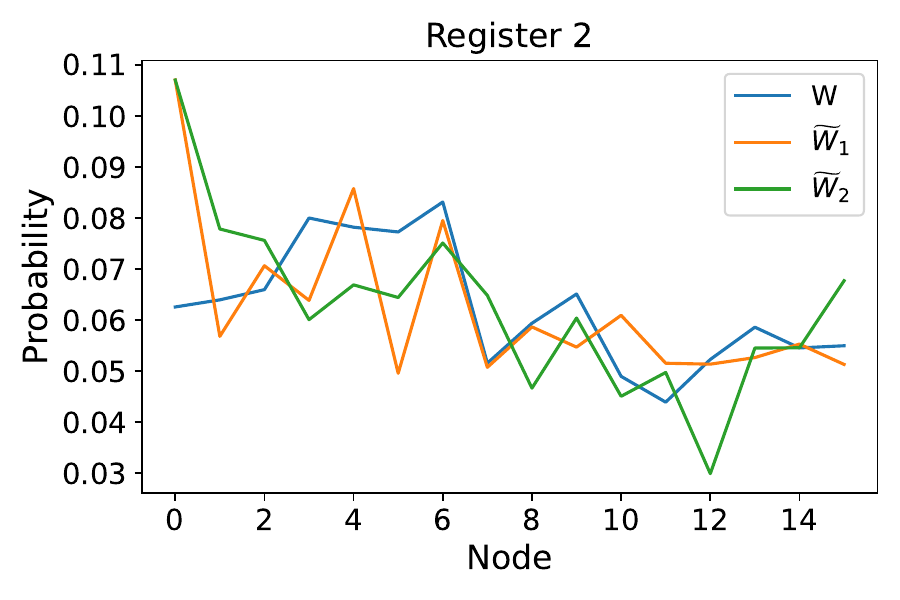}}
	\caption{Probability distributions of different Szegedy walk formulations on a random graph with $N=16$ nodes after $5$ time steps, measuring (a) the first register, and (b) the second register. The results are the same for the first register, whereas for the second register the results depend on the particular implementation of the update operator $\update$.}
	\label{F:comparison}
\end{figure*}

In this section, we analyze the effect of the implementation dependency of the update operator $\update$. First, we simulate the walks using a normalized random matrix as transition matrix $\TM$ with $N=16$ nodes. For the original walk we take the operator $W$ acting on an equal superposition of the $\left|\psi_i\right>$ states. For the similarity-transformed version, we take the operator $\widetilde{W}$ and the equal superposition of the coinless states $\left|i,0\right>$. In both cases, we perform $5$ steps and obtain probability distributions by measuring both registers. The results are shown in Figure \ref{F:comparison}. As expected by the discussion made in Section \ref{sec:Alternative}, the results for the first register are identical. However, for the second register we obtain results depending on the implementation of the update operator $\update$. Thus, this alternative formulation is only valid for algorithms measuring the first register, which indeed, is the most widespread case.

\begin{figure*}[h]
	\centering
	\subfigure[]{\includegraphics[scale=0.5]{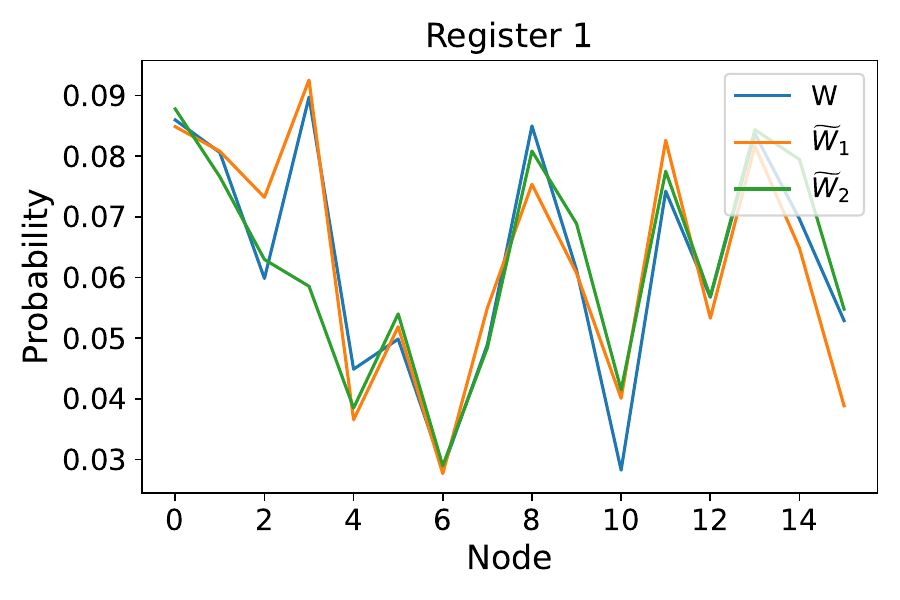}}
	\subfigure[]{\includegraphics[scale=0.5]{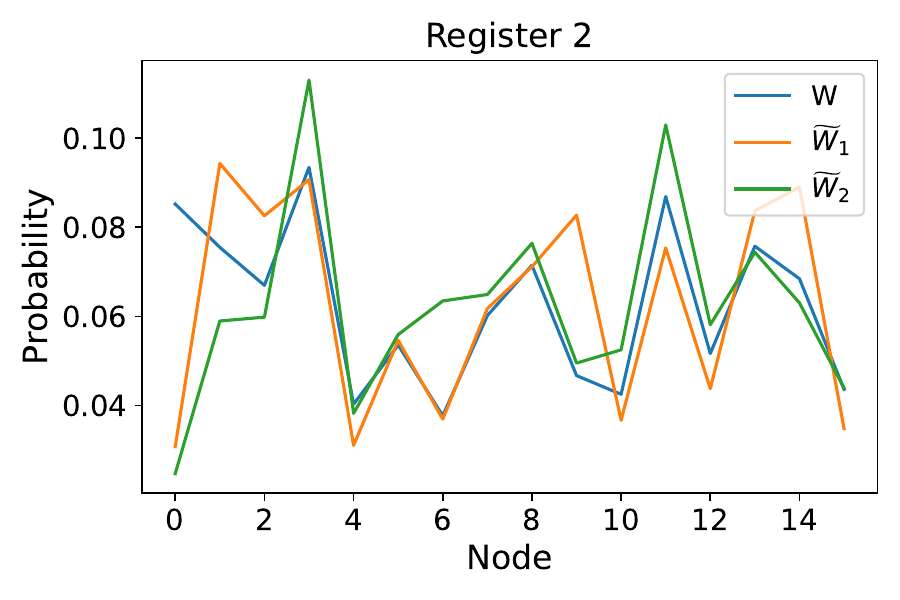}}
	\caption{Probability distributions of different Szegedy walk formulations on a random graph with $N=16$ nodes after $5$ time steps, where the walk is performed with a different Markov chain at each step. The results are obtained by measuring (a) the first register, and (b) the second register. The results for both registers are different, indicating that they depend on the particular implementation of the update operator $\update$.}
	\label{F:comparison_2}
\end{figure*}

Another important use case is when operators created from different transition matrices are concatenated. For example, let us suppose the following evolution of a state:
\begin{equation}
	\left|\phi(L)\right> = W_LW_{L-1}...W_2W_1\left|\phi(0)\right>,
\end{equation}
where $L$ different walk operators are applied. This construction is used in Ref. \cite{Lemieux} to create a heuristic algorithm for minimization based on simulated annealing using only unitary evolution. In this case, when using the alternative formulation, when two operators encounter, we have a term $V_2V_1^\dagger$. As we have said, $V^\dagger$ does not affect to the probabilities in the first register, but affects to the quantum state, and this has a non-trivial effect since this step is intermediate and there are future swaps. The state before the final $V_1^\dagger$ is not a linear combination of the states $\left|\psi^{(1)}_i\right>$, so that the state after its action is not uniquely defined. Moreover, the first operator $V_2$ of the following step, does not find a coinless state, so that again generates something not uniquely defined. Thus, we expect to find different results for both registers at the end of the sequence. We have simulated the evolution of the walks with the two different update methods for a sequence of $5$ normalized random transition matrices. The results in Figure \ref{F:comparison_2} effectively show this non-equivalence fact.

Although this illustrates that algorithms concatenating different evolution operators depend \apriori on the implementation, these algorithms are usually analyzed in a heuristic manner \cite{Lemieux,Qfold,QMS,GWQMA,Linear}, thus proving that they work for the particular implementation provided. However, there exist also theoretical studies with results that are independent of the implementation. These algorithms mix quantum evolution with non-unitary steps, as for example QPE measurements in the QSA protocol \cite{Boixo}. Particularly, for each different step the result is the corresponding stationary distribution $\left|\widetilde{\pi}\right>$ in \eqref{pi_2}, so that it is independent of the transition matrix, and thus the action of the first update operator of the following chain is always uniquely defined. See Section \ref{sec:QSA} for further details.

\section{Quantum Phase Estimation for Szegedy Quantum Walk}\label{sec:QPE}

Whereas a lot of quantum walks algorithms are based on the simple application of the unitary operator to evolve a quantum state, it turns out that the Szegedy quantum walk possesses particular spectral properties that motivates also using quantum phase estimation. For example, the quantum phase estimation algorithm has been used with the walk for problems such as detecting and finding marked nodes graphs \cite{Notes,Santha}, sampling and optimization \cite{Boixo}, reinforcement learning \cite{Paparo3}, and studying properties of electrical networks \cite{Electric}. Although these algorithms provide promising results, these are only based on theoretical analysis of the complexity. However, since simulating the Szegedy quantum walk has been so far a difficult task, there are no actual simulations proving that they work for particular systems. Moreover, numerical simulations are essential for studying the viability of heuristic algorithms where theoretical analysis is not possible \cite{QMS,Randomized}.

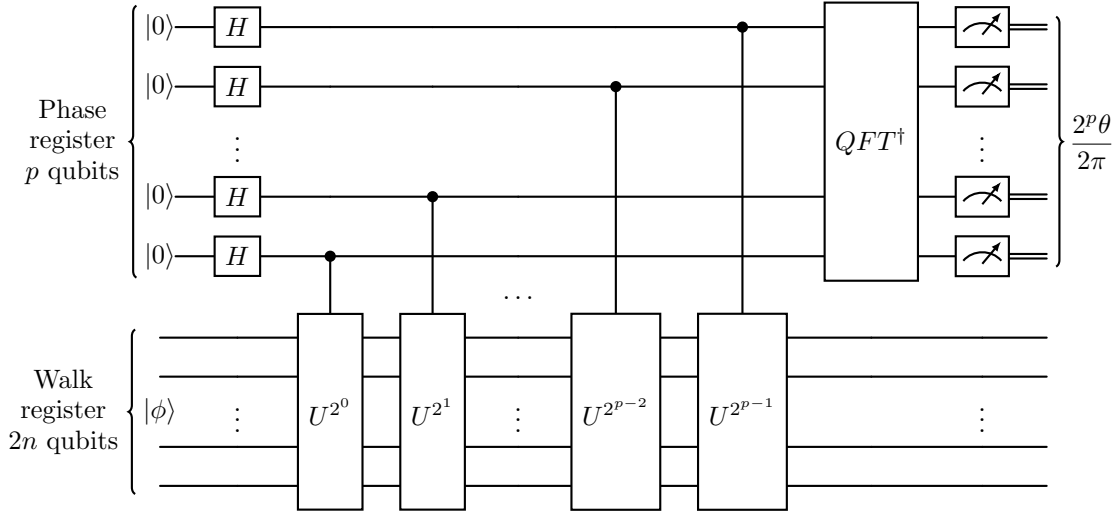
\begin{figure}[t]
	\centering
	\begin{quantikz}[row sep = 0.2cm]
		\lstick[5]{Phase\\ register\\ $p$ qubits}
		\left|0\right>&\gate{H}&&& & & \ctrl{6} &\gate[5]{QFT^\dagger}&\meter{}&\setwiretype{c}\rstick[5]{$\displaystyle\frac{2^p\theta}{2\pi}$}\\
		\left|0\right>&\gate{H}& && & \ctrl{5} &&&\meter{}&\setwiretype{c}\\
		& \vdots \setwiretype{n} &&&&&&& \vdots& \\
		\left|0\right>&\gate{H}& &  \ctrl{3} &&&& &\meter{}&\setwiretype{c}\\
		\left|0\right>&\gate{H}& \ctrl{2}  &  &  & &&&\meter{}&\setwiretype{c}\\
		& \setwiretype{n} &&& \hdots &&&& &&&\\
		\lstick[5]{Walk\\ register\\ $2n$ qubits}
		&& \gate[5]{U^{2^0}} & \gate[5]{U^{2^1}} && \gate[5]{U^{2^{p-2}}} & \gate[5]{U^{2^{p-1}}} & & &\\
		&& && &&&&&\\
		\left|\phi\right>& \vdots \setwiretype{n} &&& \vdots &&&& \vdots \\
		&& && &&&&&\\
		&& && &&&&&\\
	\end{quantikz}
	\caption{Quantum circuit for the quantum phase estimation algorithm. For a graph with $N = 2^n$, each of the two registers of the Szegedy walk requires $n$ qubits, so that the walk register as a whole requires $2n$ qubits. The unitary operator $U$ in the circuit represents in a generic form any walk evolution operator of the ones described along the paper, and the state $\left|\phi\right>$ represents an arbitrary walk state as in equation \eqref{vector}. When this is an eigenstate of $U$ with eigenvalue $e^{\ci\theta}$, the measurement in the phase register provides a $p$-bit approximation of the quantity $2^p\theta/2\pi$. If the state $\left|\phi\right>$ is a linear combination of eigenstates, then different phases can be obtained with probabilities provided by such combination.}
	\label{F:QPE}
\end{figure}

A circuit showing how the QPE algorithm works is shown in Figure \ref{F:QPE}. If an eigenvector of the walk unitary is prepared in the walk register, the result of measuring the phase register provides an estimation for the phase of the complex eigenvalue. In the case that the initial state is a linear combination of eigenstates, the phase of each of them can be measured with a probability provided by the initial amplitudes.

The naive form of simulating these algorithms would be constructing the matrix of the Szegedy walk unitary evolution operator, and then construct its equivalent controlled version for simulating the whole circuit. Nevertheless, in this work we aim to simulate the QPE algorithm using the previous algorithms for simulating the Szegedy walk, again avoiding the explicit construction of the unitary matrix. First, we show how we can simulate the action of the three main operators of the circuit---the Hadamard gates, the controlled-$U$ gates, and the quantum Fourier transform (QFT)---when they are applied to an arbitrary initial state. This would be enough to simulate any QPE-based algorithm, and necessary in the case of using the QPE unitary as an operator for a more complicated algorithm, as we will see for the search algorithm in Section \ref{sec:Search}. However, there are simpler scenarios just based on the usual QPE routine of Figure \ref{F:QPE}, i.e., starting from $\left|0\right>$ in the phase register and measuring the result. For these cases, we show a simpler direct implementation in a later section.

\subsection{Operational Implementation}

\subsubsection{State representation}

\begin{figure}[h]
	\centering
	\includegraphics[scale=0.46]{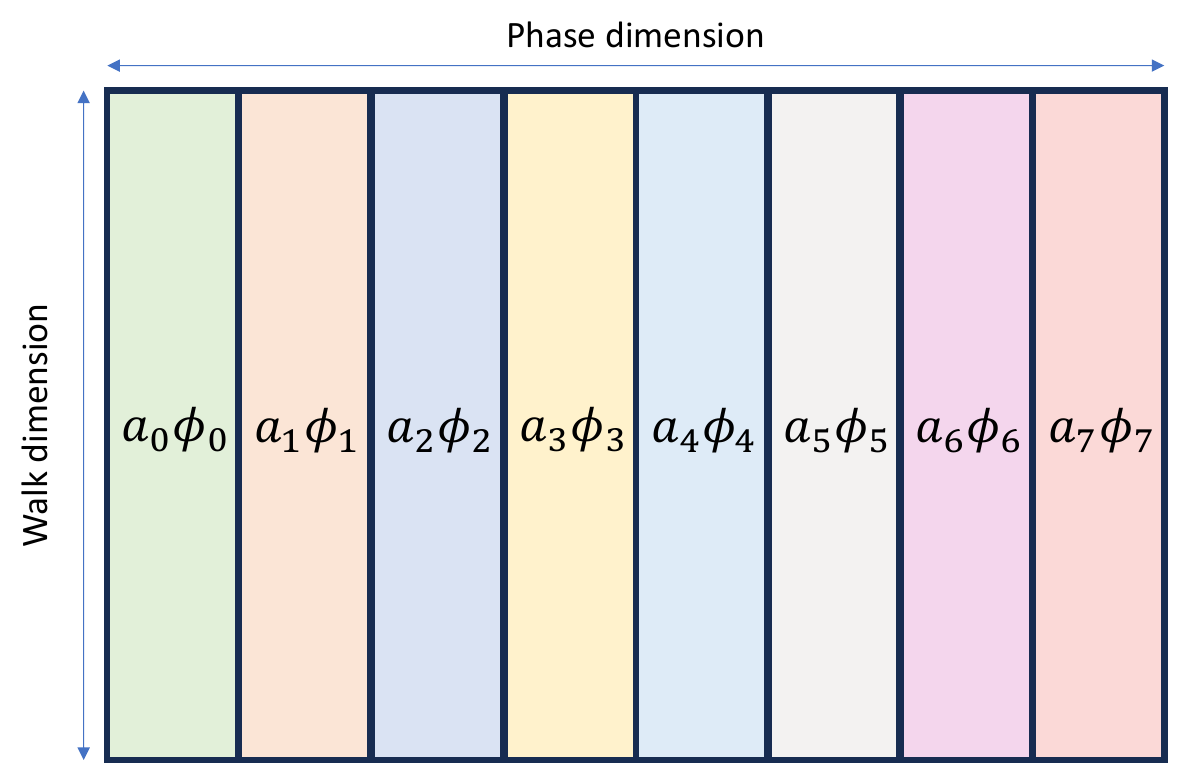}
	\caption{Matrix representation of an arbitrary state $\left|\qpestate\right>$ of the form in \eqref{PHI_qpe}. Each column corresponds to a walk state $\left|\phi_x\right>$. In the example, the phase register has $p=3$ qubits, so that there are $2^3=8$ walk states.}
	\label{F:tensor}
\end{figure}

In the context of QPE algorithms, we do not worry about the Szegedy Hilbert space being composed by two registers, and consider it as single register. Thus, in this case our states are formed by tensor products of phase states and Szegedy states. An arbitrary state can thus be expressed as:
\begin{equation}\label{PHI_qpe}
	\left|\qpestate\right> = \sum_{x=0}^{2^p-1} a_x\left|x\right>\otimes\left|\phi_x\right>,
\end{equation}
so that for each of the computational basis states $\left|x\right>$ of the phase register we have a particular Szegedy state $\left|\phi_x\right>$. Therefore, to easy the operators implementation, we represent the state as a matrix of $2^p$ columns, where each column corresponds to a Szegedy state, as shown in Figure \ref{F:tensor}. This matrix is of dimension $2^p \times N'$, where $p$ is the number of qubits in the phase register, and $N'$ the dimension of the Szegedy register. For a dense representation of the walk states, $N'=N^2$. However, if we use the sparse simulation algorithm shown in SM \cite{SM}, then $N'$ is the dimension of the reduced subspace $\mathcal{H}_R$ in \eqref{reduced}, which is reduced to $\mathcal{O}(N)$ for sparse graphs.

\subsubsection{Controlled operations}

\begin{figure}[h]
	\centering
	\makebox[10pt][c]{
	\includegraphics[scale=0.46]{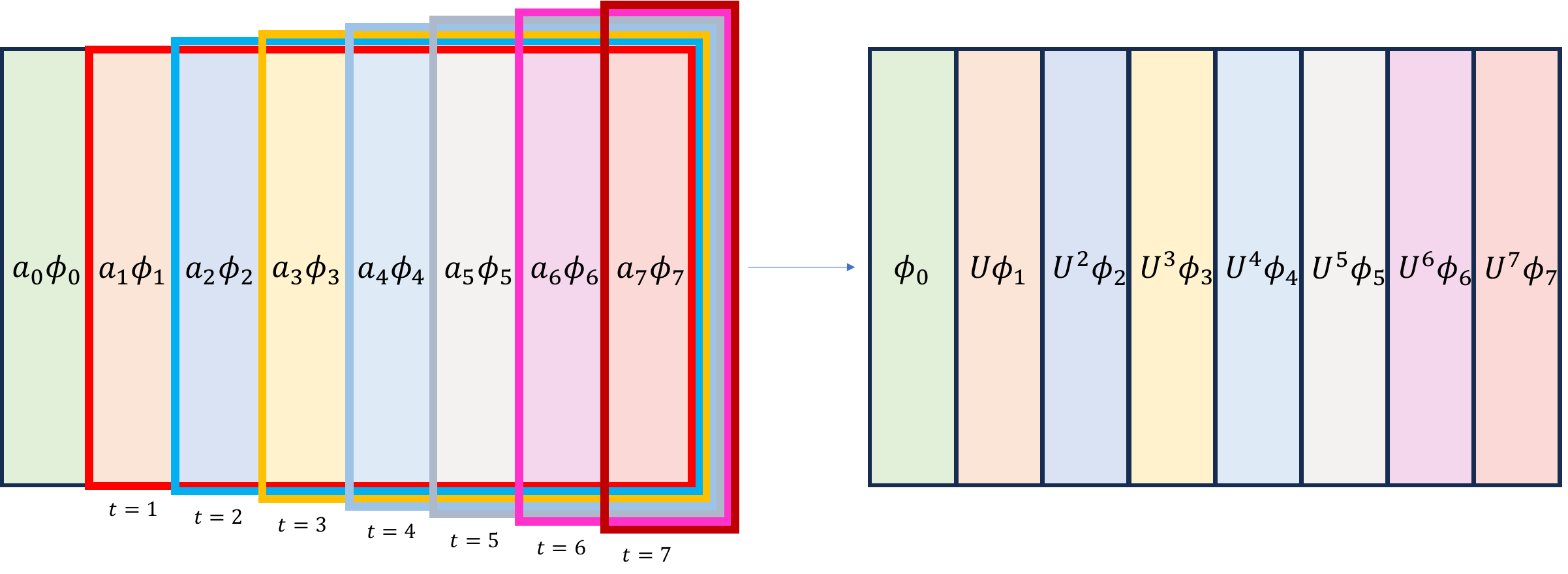}
	}
	\caption{How to apply the controlled-$U$ operations of the QPE circuit on an arbitrary state $\left|\qpestate\right>$ of the form in \eqref{PHI_qpe}, for an example with $p=3$ phase qubits. At each time step of a \textit{for} loop, the walk evolution is applied in a vectorized manner to a batch of column states. This batch decreases its size at each time step from the left, so that in the end, the $x$-th column has evolved $x$ times. The initial batch (red) comprises the states from $x=1$, decreasing its size until $x=2^p-1 = 7$ (brown).}
	\label{F:controlled}
\end{figure}

Given an arbitrary state as that of equation \eqref{PHI_qpe}, the action of the controlled-$U$ gates ($CU$) shown in Figure \ref{F:QPE}, where $U$ is an arbitrary walk unitary operator, produces the following state:
\begin{equation}\label{C_PHI_qpe}
	CU\left|\qpestate\right> = \sum_{x=0}^{2^p-1} a_x\left|x\right> \otimes U^x\left|\phi_x\right>.
\end{equation}
Therefore, to simulate this action, we need to evolve $x$ times the $x$-th column of the matrix state.  This could be performed \apriori with two \textit{for} loops, one loop running the column index, so that the unitary is applied to a single vector, and another for the $x$ times of the evolution. Nevertheless, it turns out that the memory-saving algorithms for the quantum walk simulation can act on a batch of states in a vectorized manner \cite{Squwals}. Moreover, the NumPy-based algorithm that we have devised for sparse graphs also provides this feature (see SM \cite{SM}). Note that for the first column ($x=0$), there is no evolution. Therefore, we can take a batch of states from the column $x=1$ to the last column $x=2^p-1$. After applying the evolution on this batch, the left-most vector does not need more evolution, so we remove it from the simulation batch and can proceed with the next evolution step on the remaining states. Thus, a single \textit{for} loop repeating this evolution-removal process $2^p-1$ times produces the state in \eqref{C_PHI_qpe}, since the $x$-th column remains in the size-decreasing batch only $x$ times. The whole procedure is shown in Figure \ref{F:controlled}.

\subsubsection{$H^{\otimes p}$ and quantum Fourier transform}

Both the tensor product of Hadamard gates and the QFT acting on the phase register can be applied in a similar manner. First let us show the action on an arbitrary state of the phase register:
\begin{equation}
	QFT\left(\sum_{x=0}^{2^p-1} a_x \left|x\right>\right) = \sum_{y=0}^{2^p-1}\left(\frac{1}{\sqrt{2^p}}\sum_{x=0}^{2^p-1} a_x e^{2\pi \ci xy / 2^p}\right)\left|y\right>,
\end{equation}
\begin{equation}
	H^{\otimes p}\left(\sum_{x=0}^{2^p-1} a_x \left|x\right>\right)=\sum_{y=0}^{2^p-1}\left(\frac{1}{\sqrt{2^p}}\sum_{x=0}^{2^p-1} a_x (-1)^{x \cdot y}\right)\left|y\right>.
\end{equation}
The arbitrary state in equation \eqref{PHI_qpe} can also be written as:
\begin{equation}\label{PHI_qpe_2}
	\left|\qpestate\right> = \sum_{k=0}^{N'} a_{k}\left|\varphi_k\right>\otimes\left|k\right>,
\end{equation}
where the states $\left|k\right>$ form the computational basis of the walk space where we are working. Therefore, each of the rows of the matrix state shown in Figure \ref{F:tensor} can be identified with a phase state $\left|\varphi_k\right>$. Thus, to simulate the action of these operators we have to apply the action on the $N'$ rows in parallel. However, note that the action corresponds just to the classical discrete Fourier transform (DFT)\footnote{Given the convention for the sign in the exponential on the classical DFT, for the QFT we must use the inverse fast Fourier transform, and the other way around for the $QFT^\dagger$.} or the Walsh–Hadamard transform (WHT). Moreover, these transformations can be applied in a vectorized manner for a batch of vectors, so thus we just have to apply this well-known operations along the column axis.

\subsection{Direct Implementation}\label{sec:Direct}

\begin{figure}[h]
	\centering
	\makebox[10pt][c]{
		\includegraphics[scale=0.46]{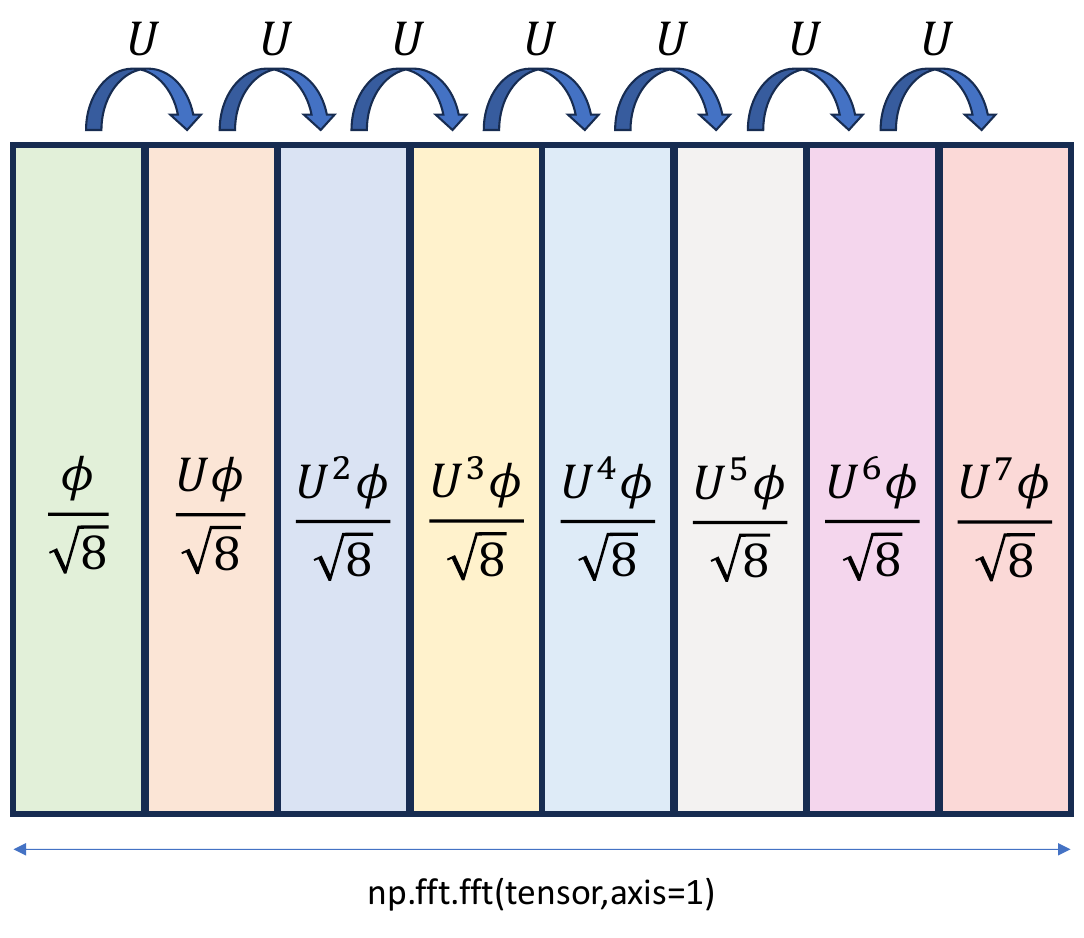}
	}
	\caption{How to obtain the state in \eqref{PHI_direct} starting from a single Szegedy state $\left|\phi\right>$, for a example with $p=3$ phase qubits. The walk unitary evolution $U$ is applied sequentially to $\left|\phi\right>$, and at each time step the result is stored in a column of the matrix state. After that, the fast Fourier transform is applied along the column axis.}
	\label{F:direct}
\end{figure}

Although the individual simulation of the operators involved in the QPE are enough, in most cases the QPE is only applied once to a state starting at $\left|0\right>$ in the phase register, so that it is simply $\left|\qpestate\right> = \left|0\right> \otimes \left|\phi\right>$ for a given Szegedy state $\left|\phi\right>$. After the application of the $H$ gates and the controlled gates we have the state
\begin{equation}\label{PHI_direct}
	(CU)H^{\otimes p}\left|\qpestate\right> = \frac{1}{\sqrt{2^p}}\sum_{x=0}^{2^p-1} \left|x\right> \otimes U^x\left|\phi\right>.
\end{equation}
Note that in this case the states in the walk register are obtained by evolving the same initial state $\left|\phi\right>$, so that if we want to form again a matrix state as in Figure \ref{F:tensor}, we just have to sequentially evolve the state $\left|\phi\right>$, and allocate the result of the $x$-th step in the $x$-column. Thus, again $2^p-1$ evolution steps are required, but now all of them are applied to a single state rather than to a batch of $\mathcal{O}(2^p)$ states. After these operations, we apply again the discrete inverse Fourier transform along the column axis to obtain the final state. The whole procedure is shown in Figure \ref{F:direct}.

\subsection{Measurements}

Depending on the algorithm, the result of interest can be provided by the phase register or by the walk register. To obtain the probability distribution of the computational basis states on the phase register, we can just take the element-wise squared modulus of an arbitrary matrix state, and add over the rows. For the walk register, if we did so for the columns, we would obtain the probability distribution for all the computational basis states of the doubled Hilbert space in \eqref{Hilbert}. However, we want the probability distribution only for one of the two registers of the walk. In this case, we first obtain the probability of the desired walk register for each of the Szegedy states $\left|\phi_x\right>$ in the matrix state, i.e., each column of our tensor. This can be done in a vectorized manner using the algorithm described in Ref. \cite{Squwals} for dense states, or the one described in SM for sparse graphs \cite{SM}. After this, we just have to add over the columns to obtain the final probability distribution of the walker.

\subsection{Multiple Phase Registers}

So far, we have seen the standard QPE algorithm where there is a single phase register. However, there are algorithms where several phase register are used to increase the precision of the protocol \cite{Santha,Electric}. Supposing we have $k$ register with $p$ qubits each, repeating the QPE operators for each phase register, the number of times the walk unitary $U$ is applied is $(2^p-1)k$, rather than $2^{kp}-1$ as would be the case if all the $kp$ qubits where used as in the usual QPE circuit. Thus, the precision is increased without introducing an exponential-growing additional time cost.

Given the tensor form of representing a quantum state, the algorithms shown previously can be generalized. In this case, we represent the state as a tensor of order $1+k$, so that there are $k$ axes indexing each phase register, and we store $2^{kp}$ Szegedy states. Thus, to perform the simulation of the operators, we just have to iteratevely applying the vectorized operations along the $i$-th axis for $i=1,...,k$.

\section{Classical Simulation with SQWLib}\label{sec:Results}

Given the simulation algorithms presented above, we have developed a Python package named SQWLib that extends the previous simulation package SQUWALS \cite{Squwals}, adding the more fundamental update $\update$ and reflection $R_0$ operators for simulating different formulations of the Szegedy quantum walk, as well as operators for simulating QPE-based walk algorithms. Moreover, we have also added methods based on NumPy for the efficient simulation of the walk on sparse graphs. This sparse methods allow the vectorization of the walk evolution on a batch of sparse states, and also the straightforward application of the DFT and WHT by aligning the states with significant zeroes (see SM for further details \cite{SM}).

In this section, we present three examples of QPE-based Szegedy walk algorithms simulated on classical computers using our package. The first one is a simple algorithm for detecting the presence of marked nodes on graphs, and we use it to show a case where the result of interest is obtained from the phase register as in a usual QPE algorithm. The other two are algorithms that leverage the spectral properties of the quantum walk for reversible Markov chains, and are used to sample distributions with great interest on optimization and machine learning algorithms.

\subsection{Detecting Marked Nodes}\label{sec:Detection}

The first algorithm that we simulate is used to detect whether there are marked nodes or not in a graph \cite{Notes}. This algorithm works when the graph satisfies two conditions: it is undirected and symmetric. Examples are the complete graph without loops and regular lattices with periodic boundary conditions. The transition matrix $\TM$ is obtained by normalizing the columns of the adjacency matrix. Then, nodes are marked transforming them into sinks, so that they only have a self-loop. Thus, the transition matrix of the marked graph is:
\begin{equation}
	\TM^{'}_{ji} =
	\left\lbrace\begin{array}{c}
		\displaystyle \TM_{ji} \ \ \ \text{if} \ i \ \notin \mathcal{M}, \\
		\displaystyle \delta_{ji} \ \ \ \ \text{if} \ i \ \in \mathcal{M},
	\end{array}\right.
\end{equation}
where $\mathcal{M}$ is the set of $M$ marked nodes.

Now suppose we can construct the initial state
\begin{equation}
	\frac{1}{\sqrt{N-M}}\sum_{i \notin \mathcal{M}}\left|\psi_i\right>,
\end{equation}
which is a uniform linear combination of the $\left|\psi_i\right>$ states corresponding to unmarked nodes. Although we do not know the marked nodes, this state could be constructed with a binary measurement that evaluates if the state in the first register is marked or not, applied to the superposition over all the nodes. Since usually $M << N$, it is very likely to obtain a negative result, projecting the state in the subspace of the unmarked nodes. In the unlikely case of a positive result, then the graph has marked nodes and we are done \cite{Notes}.

The algorithm uses QPE to detect the presence of marked nodes. In the case that $M=0$, due to the properties of the graph, this state has eigenvalue $1$, so that we always measure a phase of $0$. However, when there are marked nodes, this initial vector has only support on eigenvectors with eigenvalue different to $1$. Thus, if there are marked nodes, if the QPE has enough precision, we obtain a phase different to $0$. As stated in Ref. \cite{Notes}, this algorithm can detect the presence of marked nodes quadratically faster than the classical hitting time.

\begin{figure*}[h]
	\centering
	\subfigure[]{\includegraphics[scale=0.5]{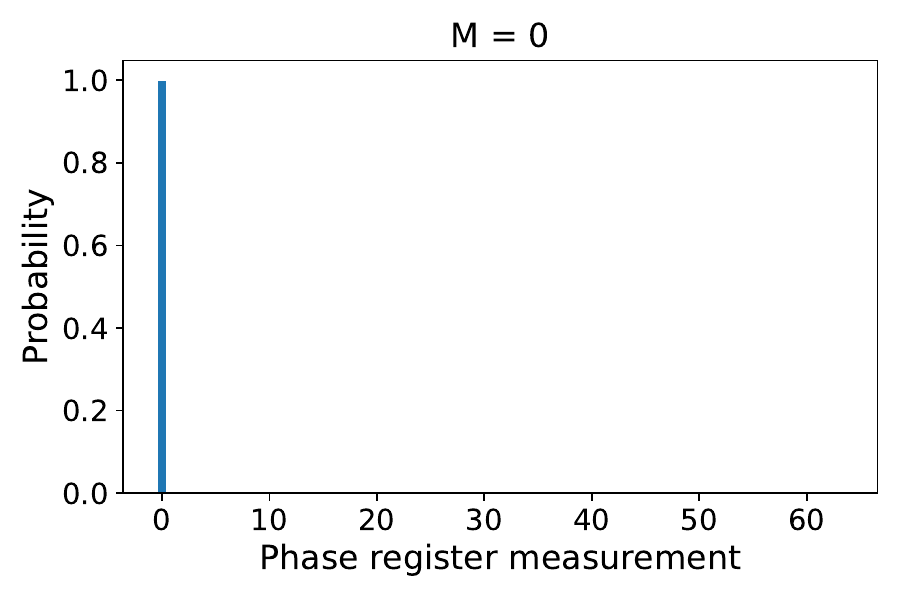}}
	\subfigure[]{\includegraphics[scale=0.5]{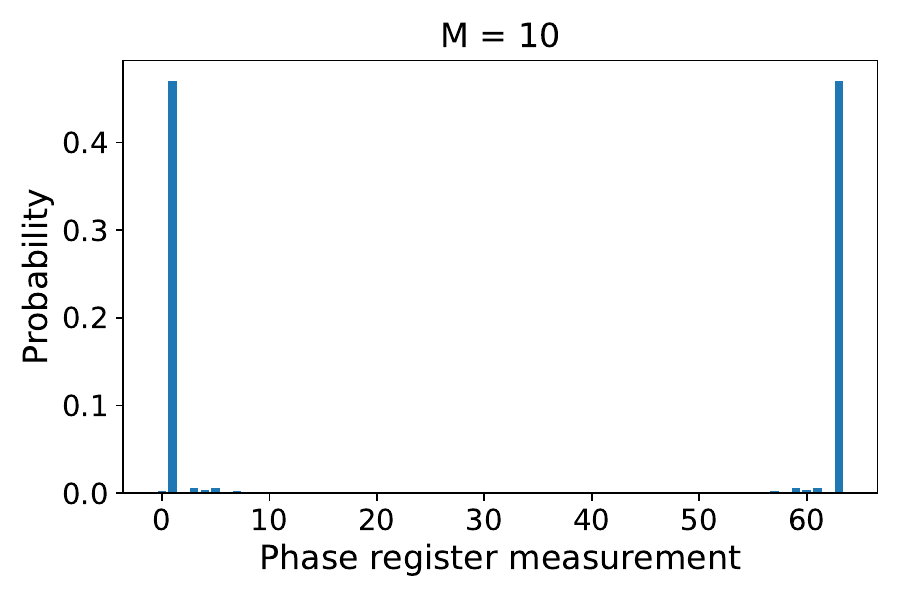}}
	\caption{Probability distribution after measuring the phase register for the QPE algorithm on a 2D lattice with $N=1024$ nodes with (a) $M=0$ marked nodes, and (b) $M=10$ marked nodes. The $x$-axis labels correspond to the decimal representation of the binary bitstring obtained by the phase register. When there are no marked nodes, the phase $0$ is always measured. However, when there are marked nodes, whatever different result is obtained instead.}
	\label{F:detect}
\end{figure*}

As an example, we have taken a 2D lattice of $N=1024$ nodes with $M=0$ and $M=10$ marked nodes, using $p=6$ qubits in the phase register. For simulating the algorithm, we use the direct implementation of the QPE explained in Section \ref{sec:Direct}. The results are shown in Figure \ref{F:detect}. When there are no marked nodes, we obtain $0$ in the phase register with probability $1$. However, when there are marked nodes, we likely obtain other phase rather than $0$. Thus, the algorithm is effectively able to detect whether the graph has marked nodes or not.

\subsection{Quantum Simulated Annealing}\label{sec:QSA}

Now, we explore the simulation of the quantum simulated annealing algorithm \cite{Boixo,Boixo_2}, which uses the spectral properties of the Szegedy quantum walk for sampling the classical stationary distribution of reversible Markov chains with speedup. This kind of chains are usually obtained from the Metropolis-Hastings algorithm \cite{Metropolis,Hastings} for Boltzmann distributions. Thus, this algorithms has high relevance in optimization problems, and also could be used to speedup machine learning algorithm as Boltzmann machines \cite{BM,Gilhan}.

In this case we use the alternative formulation with the operator $\widetilde{W}$ in \eqref{W_2}. For a reversible Markov chain, we have seen that the quantum state $\left|\widetilde{\pi}\right>$ in \eqref{pi_2}, which represents the stationary distribution of the Markov chain, is the only eigenstate with eigenvalue $1$ in the dynamical subspace. For an initial state in the this subspace, supposing so far that the QPE algorithm has infinite precision, the resulting state after the QPE protocol is of the following form:
\begin{equation}\label{qpe_state}
	a_0\left|0\right>\otimes \left|\widetilde{\pi}\right> + \sum_{i \neq 0} a_i\left|\Lambda_i\right>\otimes \left|\lambda_i\right>,
\end{equation}
where $\left|\lambda_i\right>$ are the other walk eigenvectors in the dynamical subspace, all of them with eigenvalue $\lambda$ different to $1$. Thus, after measuring the phase register, if the measurement outcome is $0$, we collapse the walk register to the stationary distribution $\left|\widetilde{\pi}\right>$. However, this occurs with a probability $|a_0|^2$, which is the overlap between the initial state and the stationary distribution, and it can be very small.

Suppose we want to sample a Boltzmann distribution, which is described as:
\begin{equation}
	p_i = \frac{\displaystyle e^{-\beta E_i}}{\displaystyle \sum_{k=0}^{N-1} e^{-\beta E_k}},
\end{equation}
where $\beta$ is a parameter regarded as inverse temperature. To sample it, the Metropolis-Hastings algorithm is used. This constructs a Markov chain whose stationary distribution corresponds to the Boltzmann distribution, whose transition matrix is:
\begin{equation}\label{MH1}
	G_{ji} =
	\left\lbrace\begin{array}{c}
		\displaystyle \frac{A_{ji}}{|B_i|} \ \text{if} \ j \in B_i,\\
		\\
		\displaystyle 1 - \sum_{k\in B_i}\frac{A_{ki}}{|B_i|}  \ \text{if} \ j = i,\\
		\\
		0 \ \text{otherwise},
	\end{array}\right.
\end{equation}
where $B_i$ is the set of nodes connected from node $i$, and the acceptance probability is defined as:
\begin{equation}\label{MH2}
	A_{ji} = \min\left(1,e^{\beta\left[E_i-E_j\right]}\right).
\end{equation}
Despite its quite complicated formulation, there is an active research on developing quantum circuits specifically designed for quantum Metropolis-Hastings walks, promising a future practical and efficient implementation on quantum computers \cite{Lemieux,Linear}.

\begin{figure}[t]
	\centering
	\makebox[10pt][c]{
	\begin{quantikz}[row sep = 0.5cm, column sep=0.4cm]
		\lstick[1]{Phase\\ register}
		\left|0\right>^{\otimes p}&\qwbundle{p}&\gate[3]{\shortstack{QPE \\ $\widetilde{W}(\beta_0)$}}&\meter{}& \ 0 \setwiretype{c}&\setwiretype{p}\left|0\right>^{\otimes p}&\gate[3]{\shortstack{QPE \\ $\widetilde{W}(\beta_1)$}}\setwiretype{q}&\meter{}& \ 0 \setwiretype{c}&\setwiretype{n}&\left|0\right>^{\otimes p}&\gate[3]{\shortstack{QPE \\ $\widetilde{W}(\beta_L)$}}\setwiretype{q}&\meter{}& \ 0 \setwiretype{c}&\setwiretype{n}\\
		& \setwiretype{n} &&&&&&&& \hdots \\
		\lstick[1]{Walk\\ register}
		\left|\widetilde{\pi}_0\right>&\qwbundle{2n}&&&&&&&&&&&\meter{} \\
	\end{quantikz}
	}
	\caption{Quantum circuit representing the quantum simulated annealing algorithm. At each time step the QPE is applied to a walk unitary $\widetilde{W}$ obtained from a Markov chain with decreasing temperature, i.e, an increasing $\beta$ parameter in the underlying Metropolis-Hastings equations \eqref{MH1} and \eqref{MH2}. The initial Szegedy state corresponds to the stationary distribution $\left|\widetilde{\pi}_0\right>$ for the first temperature $\beta_0$. Usually, $\beta_0 = 0$, so that the stationary distribution is uniform. When the measurement in the phase register provides the result $0$, the stationary distribution of the Markov chain at the corresponding temperature is obtained in the walk register. If the cooling process is slow enough, at the end the stationary probability distribution for $\beta_L$ is prepared with a probability near to $1$.}
	\label{F:QSA}
\end{figure}
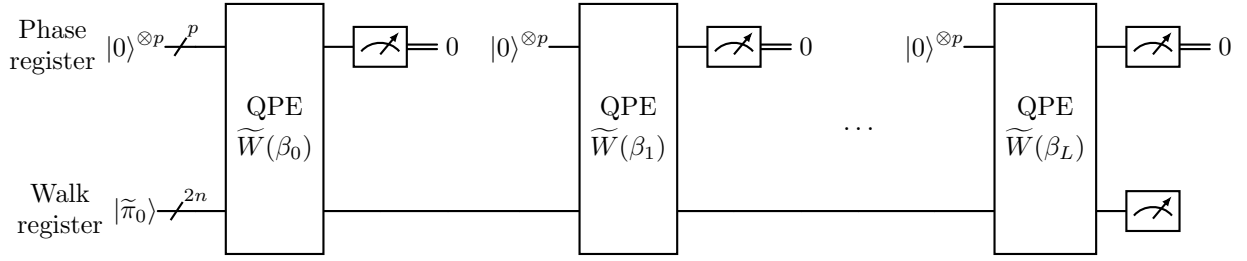

\begin{figure*}[h]
	\centering
	\subfigure[]{\includegraphics[scale=0.5]{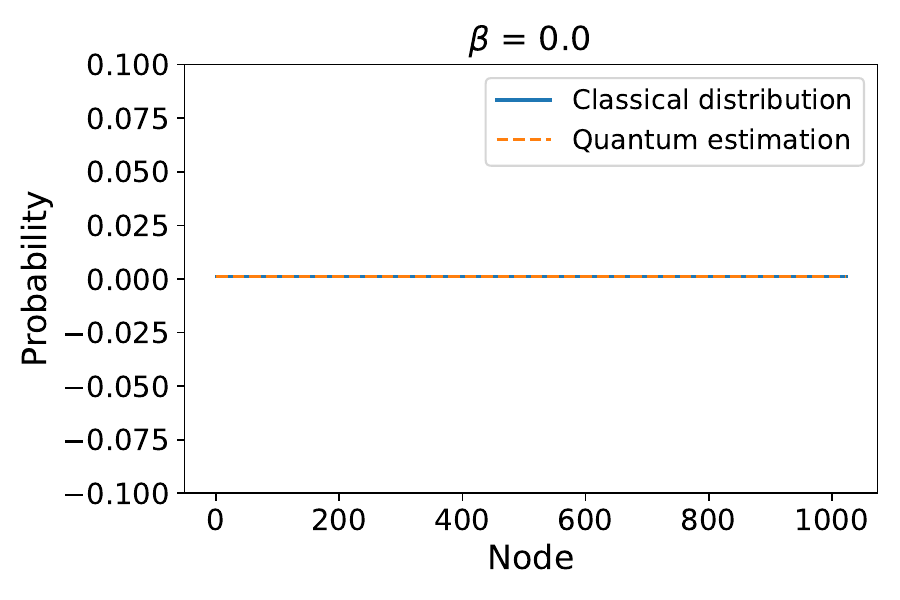}}
	\subfigure[]{\includegraphics[scale=0.5]{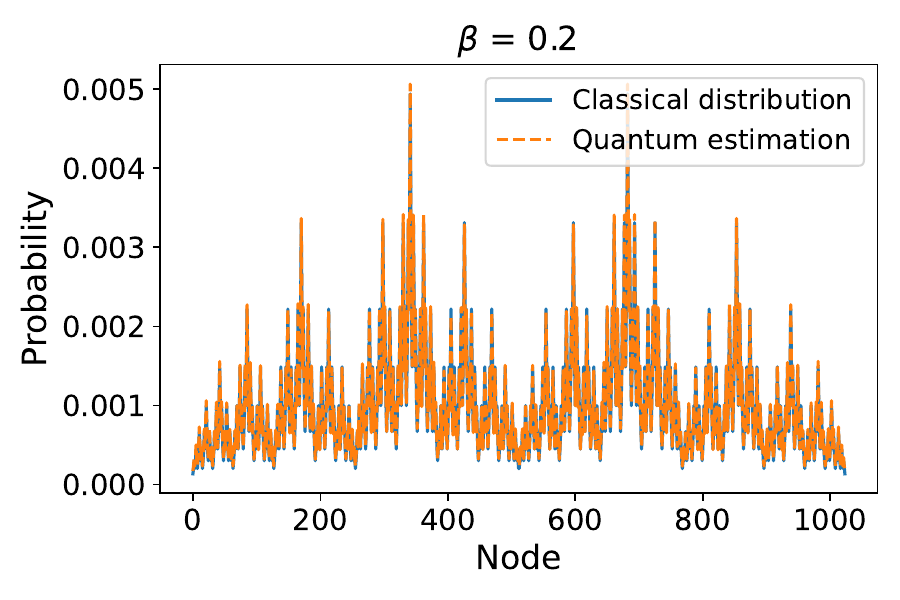}}
	\subfigure[]{\includegraphics[scale=0.5]{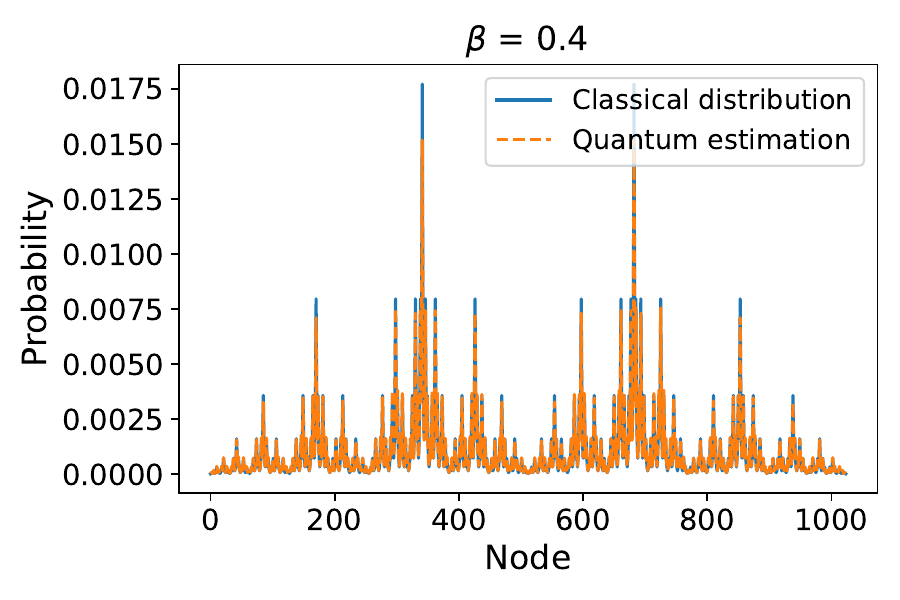}}
	\subfigure[]{\includegraphics[scale=0.5]{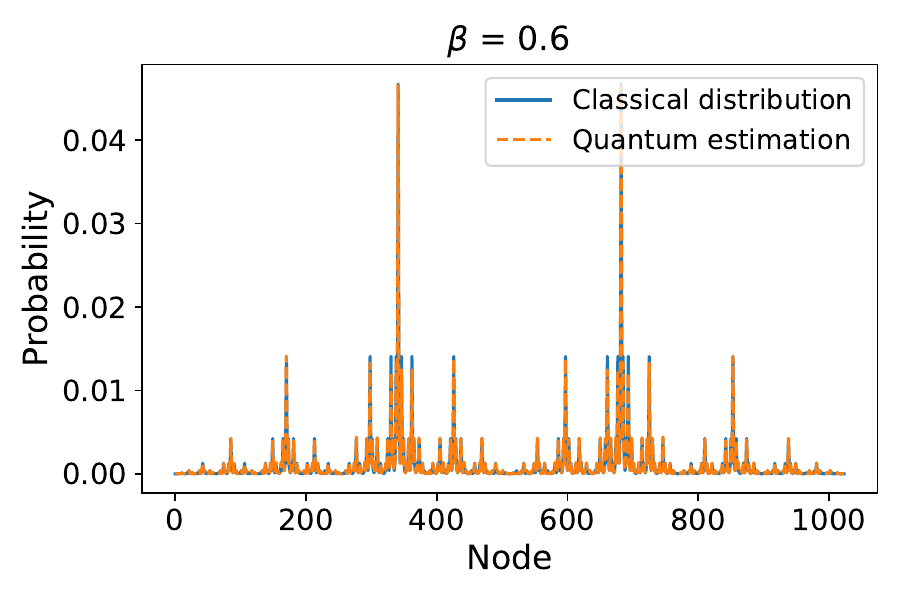}}
	\subfigure[]{\includegraphics[scale=0.5]{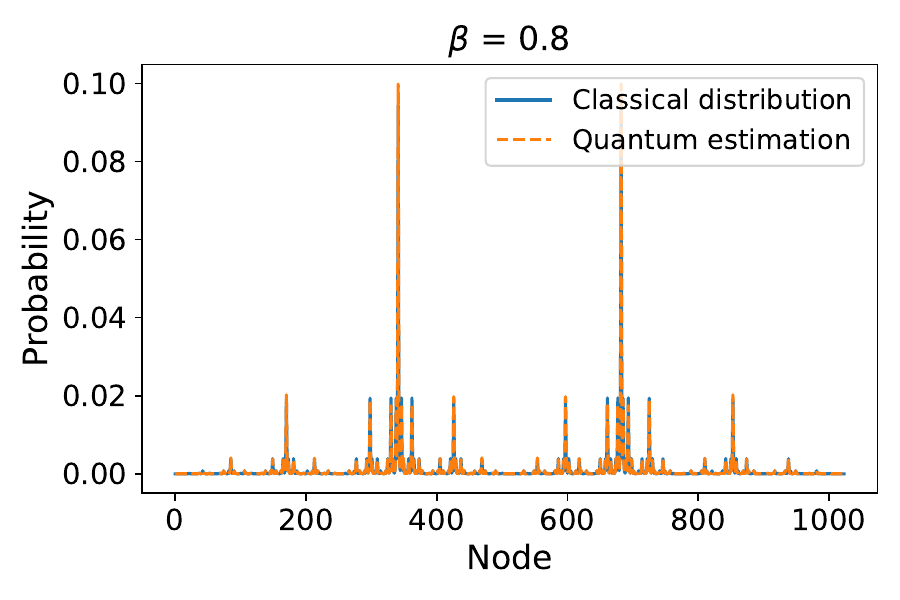}}
	\subfigure[]{\includegraphics[scale=0.5]{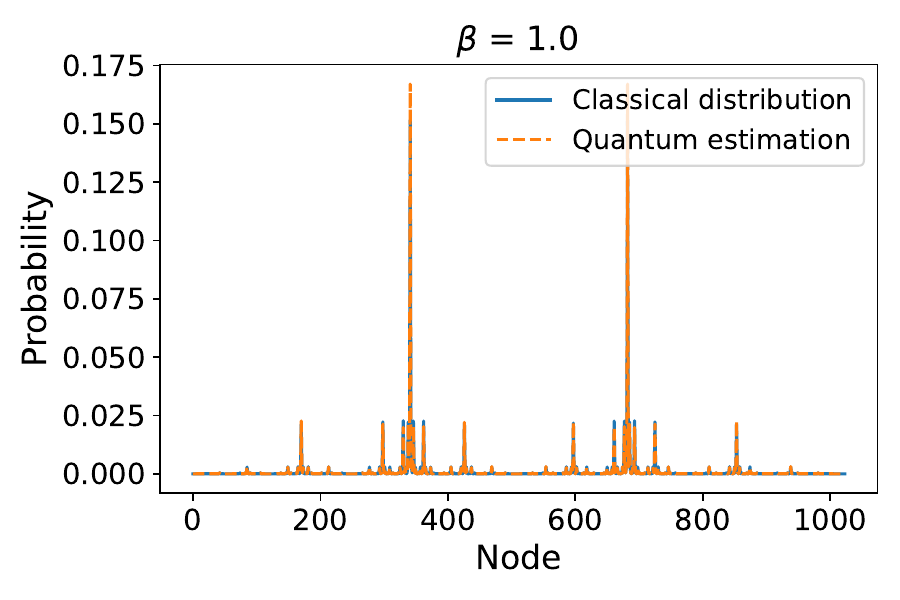}}
	\caption{Probability distributions obtained by measuring the first register of the walk for $5$ cooling steps from $\beta=0$ to $\beta=1$, in the QSA protocol applied to a Metropolis-Hastings Markov chain obtained from a Ising model with $n=10$ qubits, with $p=3$ phase qubits. At each step, the resulting quantum distribution (orange) overlaps extensively with the corresponding classical stationary distribution (blue).}
	\label{F:QSA_1}
\end{figure*}

To sample the Boltzmann distribution for a particular value of $\beta$, an annealing schedule is proposed \cite{Boixo}. The system starts with a Markov chain whose stationary distribution can be easily prepared, for example the uniform one at infinite temperature, and the system is cooled slowly until the desired temperature, obtaining the stationary distribution at each step. Thus, if the cooling is slow enough, by Zeno effect, the probability of measuring phase $0$ at each intermediate step is almost $1$, and at the end we obtain the desired state with high probability. An schematic example of the protocol in terms of circuits is shown in Figure \ref{F:QSA}. Taking into account the finite precision of the QPE algorithm, it was found a quadratic speedup in the number of walk steps for sampling the stationary distribution with respect to the classical mixing time. See Ref. \cite{Boixo} for further details.

Note that in this case we need to use the alternative formulation, so that the quantum stationary state is independent of the transition matrix, and thus the result for one step is in the dynamical subspace of the following step. If this would not be the case, the state would have support outside the dynamical subspace, and thus support on other eigenvalues with eigenvalue $1$ apart from the stationary distribution. Therefore, the stationary distribution would not be perfectly obtained even assuming infinite precision in the QPE protocol.

In this work we show a simulation for obtaining the Boltzmann distribution of a linear spin-glass system with $n=10$ qubits. For simplicity, we only allow transitions between spin states related by a single bit-flip, so that each node of the $2^{10}=1024$ nodes of the graph has only $n$ neighbors. Thus, it is sparse, and we can simulate it more efficiently with our sparse implementation of the walk. For the Ising spin glass we take the following energy function:
\begin{equation}
	E = J \sum_{i=1}^{n-1} s_i s_{i+1},
\end{equation}
where given a bitstring, $s=-1$ if the bit is $0$, and $+1$ if the bit has value $1$. We take $J=1$ for the coupling constant. Regarding the phase register, we have taken $p=3$ qubits. Our objective is to sample the Boltzmann distribution for $\beta = 1$, and first we show an example using $5$ cooling steps from $\beta = 0$, starting from the uniform distribution. So far, at each step of the cooling process we assume that the result of the phase measurement is always $0$, so that we take the first column of our tensor state and normalize it to obtain the resulting walk state. This is used to calculate the probability distribution of the walker, and as initial state for the QPE of the following step. Note that since at each step the phase register starts at $\left|0\right>$, we can use the direct implementation of the QPE protocol. In Figure \ref{F:QSA_1} we show the stationary distribution that is prepared by the QPE subroutine at each cooling step, observing that it overlaps with the classical one.

\begin{figure*}[h]
	\centering
	\subfigure[]{\includegraphics[scale=0.5]{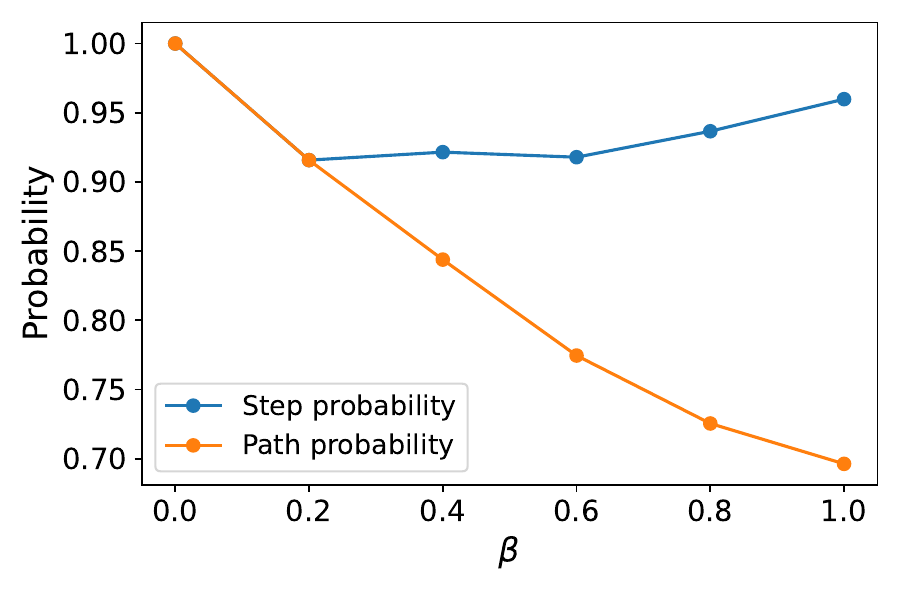}\label{F:QSA_2_a}}
	\subfigure[]{\includegraphics[scale=0.5]{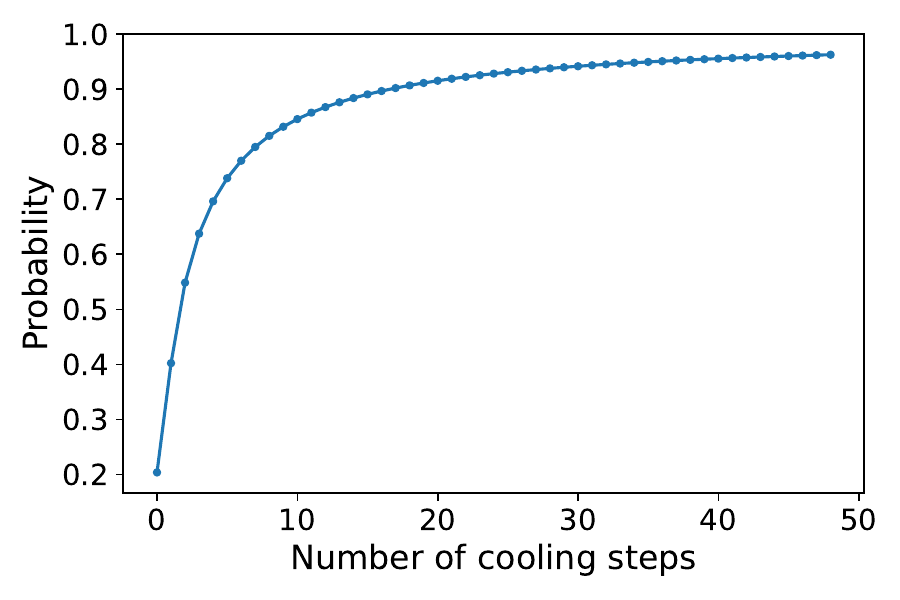}\label{F:QSA_2_b}}
	\caption{(a) Probability of measuring phase $0$ at each cooling step for the QSA protocol with $5$ cooling steps from $\beta=0$ to $\beta=1$, applied to a Metropolis-Hastings Markov chain obtained from a Ising model with $n=10$ qubits, with $p=3$ phase qubits. It is shown the individual probability at each time step, as well as the joint probability of obtaining $0$ at all the previous steps. (b) Probability of measuring phase $0$ at all the cooling steps of the QSA protocol, for different simulations with increasing number of cooling steps. For a slow enough cooling process, the total probability is near to $1$.}
	\label{F:..}
\end{figure*}

Although the results for the distributions are good, we also need to analyze what is the real probability of measuring $0$ at each cooling step. In Figure \ref{F:QSA_2_a} we show the probability of measuring $0$ at each step assuming that all previous steps have also resulted in $0$, and also the product probability, which indicates the probability of all the measurement steps so far resulting in $0$. We can observe that the probability at each step is always above $0.9$. However, the probability that all the steps produce $0$ drops to $0.7$. This is so because we have few cooling steps, so that is not slow enough. Thus, to study the effect of the number of cooling steps, we have simulated the algorithm for increasing number of steps from $\beta = 0$ to $\beta=1$. The results in Figure \ref{F:QSA_2_b} show that for enough cooling divisions, the probability of obtaining at the end the Boltzmann distribution can be increased near to the unity, so that the QSA can be effectively used for sampling the classical stationary distribution. Although the results have been obtained for one of the two simulation methods of the update operator $\update$, we have checked that they are identical for both update operators. Thus, this algorithm does not depend on the implementation, as expected.

Finally, we want to mention that after the formulation of the QSA that we have shown here, another different one proposing a randomized algorithm that avoids the QPE protocol was proposed \cite{Boixo_2}. However, it can only be used to take samples of the stationary distribution, but it is not able to prepare it as a quantum state if it is not coupled with a QPE protocol at the end. Although just sampling is by itself of great importance, for example for Boltzmann machines, preparing the quantum state has also great importance for other algorithms, as for example the search on graphs shown in the next section. Moreover, different algorithms for quantum simulated annealing have been proposed improving the complexity, also using QPE \cite{Wocjan}.

\subsection{Quantum Search on Graphs}\label{sec:Search}

The previous algorithms whose simulation we have shown use the QPE in the usual manner, with the phase register starting in state $\left|0\right>$ and measuring at the end. Thus, they could be simulated using the direct implementation that we described in Section \ref{sec:Direct}. However, there are algorithms where the QPE unitary is used as an operator inside more complex circuits, so that we have to simulate the action on arbitrary states with the operational framework. An example is the quantum walk search on graphs \cite{Santha}. The aim of this algorithm is to search for nodes marked by an oracle, performing the amplitude amplification algorithm from the quantum stationary distribution $\left|\pi\right>$ in \eqref{pi}. At the moment of proposing this algorithm, the ability to construct this initial state was assumed as efficient. However, note that it indeed can be realized with the QSA algorithm of the previous section.

The amplitude amplification algorithm is a variant of the Grover algorithm \cite{Grover,Grover2,Grover_M} that allows starting with an arbitrary initial state. Let us express the initial state in the computational basis as:
\begin{equation}
	\sum_{i=0}^{N-1}a_i\left|i\right>.
\end{equation}
The goal of the algorithm is to produce a state near to a linear combination of the marked states:
\begin{equation}
	\frac{\displaystyle \sum_{i \in \mathcal{M}}a_i\left|i\right>}{\displaystyle \sqrt{\sum_{i \in \mathcal{M}}|a_i|^2}}.
\end{equation}
Note that when performing this algorithm, not only a marked state is found, but it is also sampled with the relative probability in the initial state. Thus, when the initial state is the stationary distribution of the Markov chain, the algorithm provides samples of marked states according to it. This has great relevance in the decision of classical agents in reinforcement learning, since this process consists on deciding an action (marked node) with a probability taken from the stationary distribution of a Markov chain. For this reason, a quantum algorithm based on this search was proposed for speeding up classical reinforcement learning tasks \cite{Paparo3}.

The amplitude amplification protocol requires two operators. One is an oracle for deciding whether a state satisfy a certain condition, i.e., it is marked. The other is a reflection operator around the initial state. Whereas in the Grover algorithm the reflection is trivial due to the form of the initial state, for the stationary distribution of a Markov chain it is a real challenge, since it is \apriori unknown. Therefore, a method using QPE was proposed to approximate it.

\begin{figure}[h]
	\centering
	\resizebox{\columnwidth}{!}{%
		\begin{quantikz}[column sep = 0.05cm,row sep = 0.1cm]
			\lstick[7]{Phase\\ registers\\ $kp$ qubits}
			&\gate{H}&&& & &&& \ctrl{8} &\gate[7]{QFT^\dagger}&\octrl{2}&\gate[7]{QFT}&&&&&&&\ctrl{8}&\gate{H}&\\
			&\setwiretype{n}&& \vdots \setwiretype{n} &&&&&&&&&&&&& \\
			&\gate{H}& && && \ctrl{7} &&&&\octrl{2}&&&&&&\ctrl{7}&&&\gate{H}&\\[0.4cm]
			& \vdots \setwiretype{n} &&&&&&& &&&&&&&&&&&\vdots&&&&\\[0.4cm]
			&\gate{H}& &  &\ctrl{4} &&&&&&\octrl{2}&&&&\ctrl{4} &&&&&\gate{H}&\\
			&\setwiretype{n}&& \vdots \setwiretype{n} &&&&&&&& \\
			&\gate{H}& \ctrl{2}  &  &&&&&&&\octrl{0}&&\ctrl{2}  & &&&&&&\gate{H}&\\
			& \setwiretype{n} && \hdots &&\hdots&&\hdots&&&&&& \hdots &&\hdots&&\hdots&\\
			\lstick[5]{Walk\\ register\\ $2n$ qubits}
			&& \gate[5]{W^{2^1}} && \gate[5]{W^{2^{p-1}}} &&\gate[5]{W^{2^1}} && \gate[5]{W^{2^{p-1}}} &&&& \gate[5]{\left(W^\dagger\right)^{2^1}} && \gate[5]{\left(W^\dagger\right)^{2^{p-1}}} &&\gate[5]{\left(W^\dagger\right)^{2^1}} && \gate[5]{\left(W^\dagger\right)^{2^{p-1}}}&&\\
			&& && &&&&&&&&&&&&&&&&\\
			& \vdots \setwiretype{n} &&& \vdots &&&&&&&&&&&&&&& \vdots \\
			&& && &&&&&&&&&&&&&&&&\\
			&& && &&&&&&&&&&&&&&&&\\
		\end{quantikz}
	}
	\caption{Quantum circuit for the approximate reflection $R_s$ around the stationary state $\left|\pi\right>$ in the dynamical subspace $\mathcal{H}_D$. There are $k$ phase register, with $p$ qubits each. First, the QPE unitary subroutine is applied. Then, a reflection around the state $\left|0\right>$ in the phase register is performed (up to a global phase) with a multi-controlled-phase gate. Finally, the QPE is uncomputed.}
	\label{F:reflection}
\end{figure}
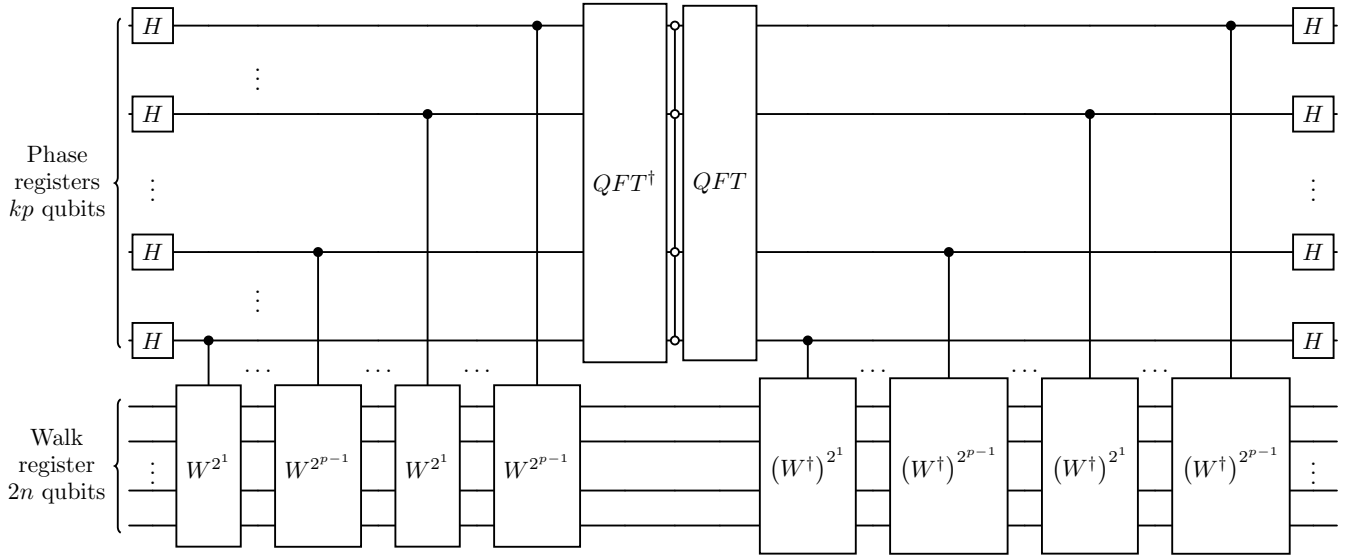

This time we use the original formulation of the walk $W$ in \eqref{W}. For a reversible Markov chain, the state $\left|\pi\right>$ in \eqref{pi} is the only eigenstate with eigenvalue $1$ in the dynamical subspace. Moreover, the application of the oracle keeps the system in this subspace. Thus, if we perform the QPE unitary protocol, without measuring, for an arbitrary intermediate state of the algorithm, we again obtain a state similar to the one shown in \eqref{qpe_state}, where only the stationary distribution has in the phase register associated the state $\left|0\right>$. The idea is then to reflect around the state $\left|0\right>$ in the phase register, so that all the other states of the combination flip their sign, and uncompute the QPE unitary. Since the state is always in the dynamical subspace, the total action is a reflection only around the stationary distribution state $\left|\pi\right>$. Finally, to reduce the effects of the finite precision of the QPE protocol, the QPE is repeated with several phase register, and the reflection in the phase register is such that flips the sign of whichever state that has a phase different to $0$ in any phase register. Thus, it corresponds again to a reflection around the state $\left|0\right>$ of all the phase register considered as a single one. A quantum circuit showing the approximate reflection operator is shown in Figure \ref{F:reflection}.

\begin{figure*}[h]
	\centering
	\subfigure[]{\includegraphics[scale=0.5]{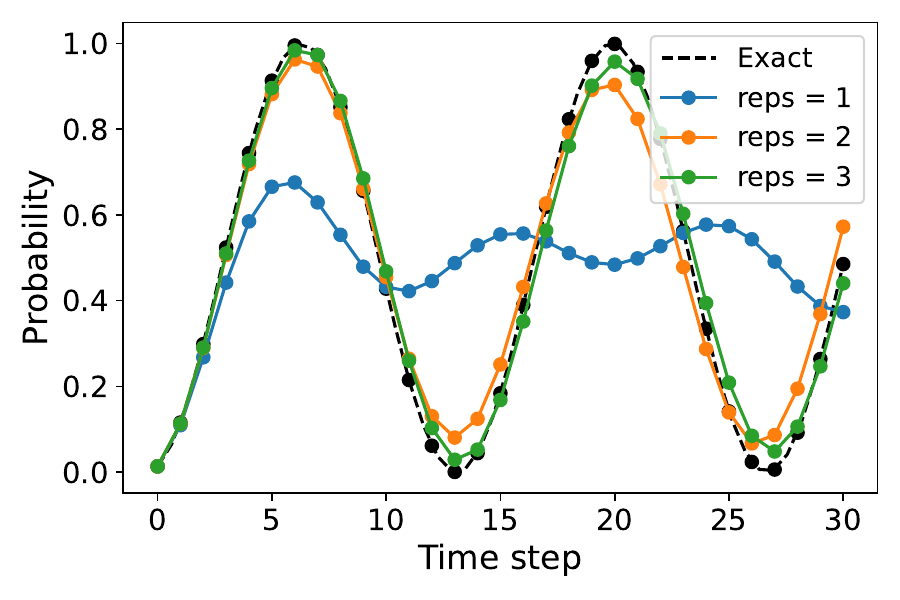}\label{F:search}}
	\subfigure[]{\includegraphics[scale=0.5]{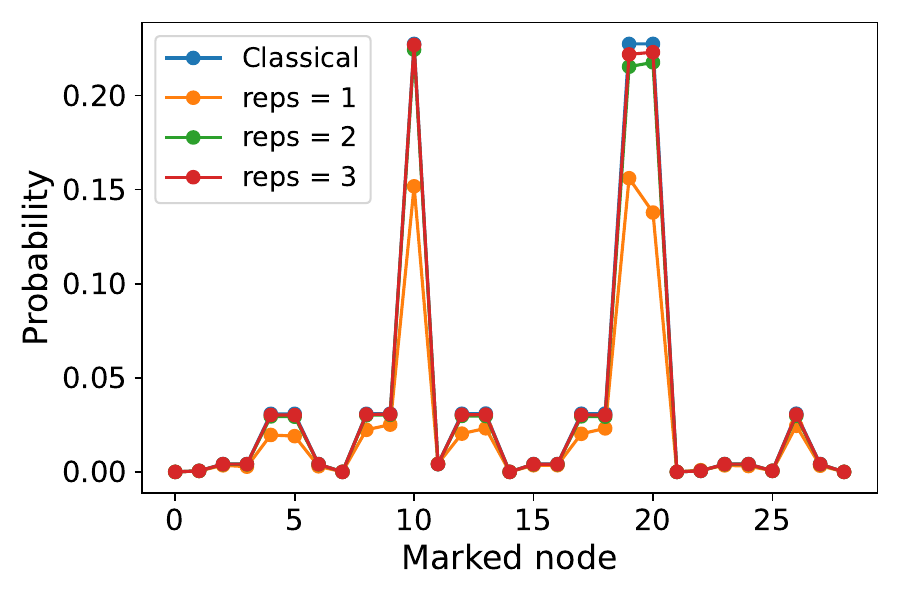}\label{F:search_2}}
	\caption{Results for the simulation of the QPE-based quantum walk search on graphs algorithm applied to a graph with $N=1024$ nodes and $M=29$ marked nodes, with increasing number of phase registers, with $p=3$ qubits each. (a) Probability of measuring a marked node in the first register of the Szegedy walk. As the number of phase registers increases, the curve approaches the theoretical one for a perfect amplitude amplification protocol. (b) Probability distribution obtained at the first maximum of probability of the marked nodes. For three phase registers, the quantum distribution practically overlaps with the classical stationary distribution on the marked nodes.}
	\label{F:...}
\end{figure*}

In our work, we simulate the naive form of the amplitude amplification algorithm for a reversible Markov chain of $N=1024$ nodes with $M=29$ marked nodes. Indeed, it is the same chain used for the QSA in the previous section. The algorithm consists on repeatedly applying the operator $R_sQ_1$, where $R_s$ is the reflection around the stationary distribution described above, and $Q_1$ is an oracle acting on the first register of the walk. To simulate the oracle $Q_1$ we can use the algorithm in Ref. \cite{Squwals} for dense graphs, or the one described in SM for sparse graphs \cite{SM}. As for the reflection around $0$ in the phase register, it can be simulated up to a global phase just multiplying by $-1$ the vector of our tensor state corresponding to index $0$ for all the phase axes. The results of the probability of measuring a marked node for each time step are shown in Figure \ref{F:search}, using $p=3$ phase qubits and different repetitions of the phase register. We can observe that as we increase the number of repetitions, the curve reaches the analytical results for a perfect amplitude amplification. If we measure at the first maximum time, we sample the stationary distribution supported only by the marked nodes, as shown in Figure \ref{F:search_2}.

\section{Conclusions}\label{sec:Conclusions}

In this work, we developed an efficient classical simulation framework for Szegedy quantum walks and related phase-estimation-based algorithms. By formulating the walk in terms of the more fundamental update and reflection operators, we enabled the simulation of a broader class of Szegedy walk constructions beyond the original formulation. Since the update operator is not uniquely defined, we also introduced two different implementations and used them to identify when the resulting dynamics are, and are not, independent of the particular implementation choice.

We further developed classical simulation methods for the operators appearing in phase-estimation-based walk algorithms, again avoiding the explicit construction of full unitary matrices. This makes it possible to study algorithms that go beyond direct walk evolution while preserving the efficiency of the underlying simulation framework. Our methods also accommodate multiple phase registers, thereby extending the framework to more general phase-estimation-based constructions.

These methods were implemented in the Python package SQWLib, including a sparse NumPy-based implementation that enables efficient simulations on large sparse graphs. Using this framework, we demonstrated representative simulations of marked-node detection, quantum simulated annealing, and quantum search on graphs. To the best of our knowledge, this constitutes the first numerical simulation framework for such phase-estimation-based Szegedy-walk algorithms beyond purely theoretical analysis.

Overall, this work provides a practical and efficient framework for the numerical study of a broad class of Szegedy-walk-based algorithms. In future work, it would be interesting to apply these methods to more complete end-to-end settings, such as Boltzmann machines or reinforcement learning agents, in order to assess their practical behavior and infer their effective complexity through simulation. More broadly, the QPE simulation framework developed here is not restricted to Szegedy walks and could be extended to other quantum algorithms of interest.

\section{Data Availability Statement}\label{Data}

The simulator library, as well as the codes for the simulation results of this paper, are available on GitHub: \url{https://github.com/qDNA-yonsei/SQWLib}.

\section*{Acknowledgments}

This work is supported by Institute of Information \& communications Technology Planning \& evaluation (IITP) grant funded by the Korea government (No. 2019-0-00003, Research and Development of Core Technologies for Programming, Running, Implementing and Validating of Fault-Tolerant Quantum Computing System), the National Research Foundation of Korea (RS-2025-02309510), the Ministry of Trade, Industry, and Energy (MOTIE), Korea, under the Industrial Innovation Infrastructure Development Project (RS-2024-00466693), and by Korean ARPA-H Project through the Korea Health Industry Development Institute (KHIDI), funded by the Ministry of Health \& Welfare, Korea (RS-2025-25456722).

\bibliography{MiBiblio}
\bibliographystyle{unsrt}

\onecolumngrid
\newpage
\begin{center}
	\textbf{\large Supplementary Material: Efficient Simulation of Szegedy Quantum Walk Formulations and Algorithms}
\end{center}
\setcounter{equation}{0}
\setcounter{section}{0}
\setcounter{figure}{0}
\setcounter{table}{0}
\setcounter{page}{1}
\makeatletter
\renewcommand{\theequation}{S\arabic{equation}}
\renewcommand{\thesection}{S\Roman{section}}
\renewcommand{\thefigure}{S\arabic{figure}}
\renewcommand{\bibnumfmt}[1]{[S#1]}
\renewcommand{\citenumfont}[1]{S#1}

\begin{center}
	Sergio A. Ortega$^1$ and Daniel K. Park$^{1,2,3}$
	
	$^1$\textit{Department of Statistics and Data Science, Yonsei University, Seoul 03722, Republic of Korea}
	
	$^2$\textit{Department of Applied Statistics, Yonsei University, Seoul 03722, Republic of Korea}
	
	$^3$\textit{Department of Quantum Information, Yonsei University, Seoul 03722, Republic of Korea}
\end{center}

\section{Quantum Walk Sparse Simulation}

Although the Szegedy quantum walk is defined in a Hilbert $\mathcal{H}$ space of dimension $N^2$ for a graph with $N$ nodes, it turns out that all the dynamics can be reduced to a reduced subspace $\mathcal{H}_R$ defined as:
\begin{equation}\label{SM_reduced}
	\mathcal{H}_R := \text{span}\lbrace{\left|i\right>_1\left|j\right>_2 : A_{ji} = 1\rbrace},
\end{equation}
where $A$ is the adjacency of the underlying undirected graph. This subspace is formed by all the edges whose transition weight is non-zero, as well as their swapped versions. Therefore, for sparse graphs the dimension of the reduced subspace is $N_R << N^2$, so that the quantum state vectors are also sparse. Usually, sparse graphs have $\mathcal{O}(N)$ edges, and thus the reduced dimension scales linearly with the number of nodes $N$. Taking this into account, we could save both time and memory resources using a sparse representation of the mathematical objects involved in the simulation.

\Apriori we could implement the simulation algorithms showed in the main text with a package able to manage sparse matrices like SciPy. This would be enough if we just wanted to simulate the quantum walk evolution on a single state. However, we are interested also in applying the discrete Fourier and Hadamard transforms along several states, which is not straightforward for usual sparse objects, since the non-null elements are not aligned. For this reason, we show a method for simulating the quantum walk storing only the elements in the reduced subspace $\mathcal{H}_R$, working directly with NumPy. In contrast to usual sparse representations, which only store non-null elements along with their matrix indexes, here we store all the elements in $\mathcal{H}_R$ including zeroes in a single vector. Doing so, their indexes are determined by the position in such vector, avoiding the need to store the indexes separately for each object. Thus, when we have different states stacked as in Figure 5 of the main text, the relevant elements of such states are aligned and we can easily apply the mentioned transformations. Moreover, our NumPy-based methods allow the parallelization of the evolution on several states, which is also a desired requirement for the simulation of the controlled gates in the quantum phase estimation algorithm.

\subsection{Reduced Subspace and Sparse Representation}

\begin{figure}[h]
	\centering
	\makebox[10pt][c]{
	\includegraphics[scale=0.4]{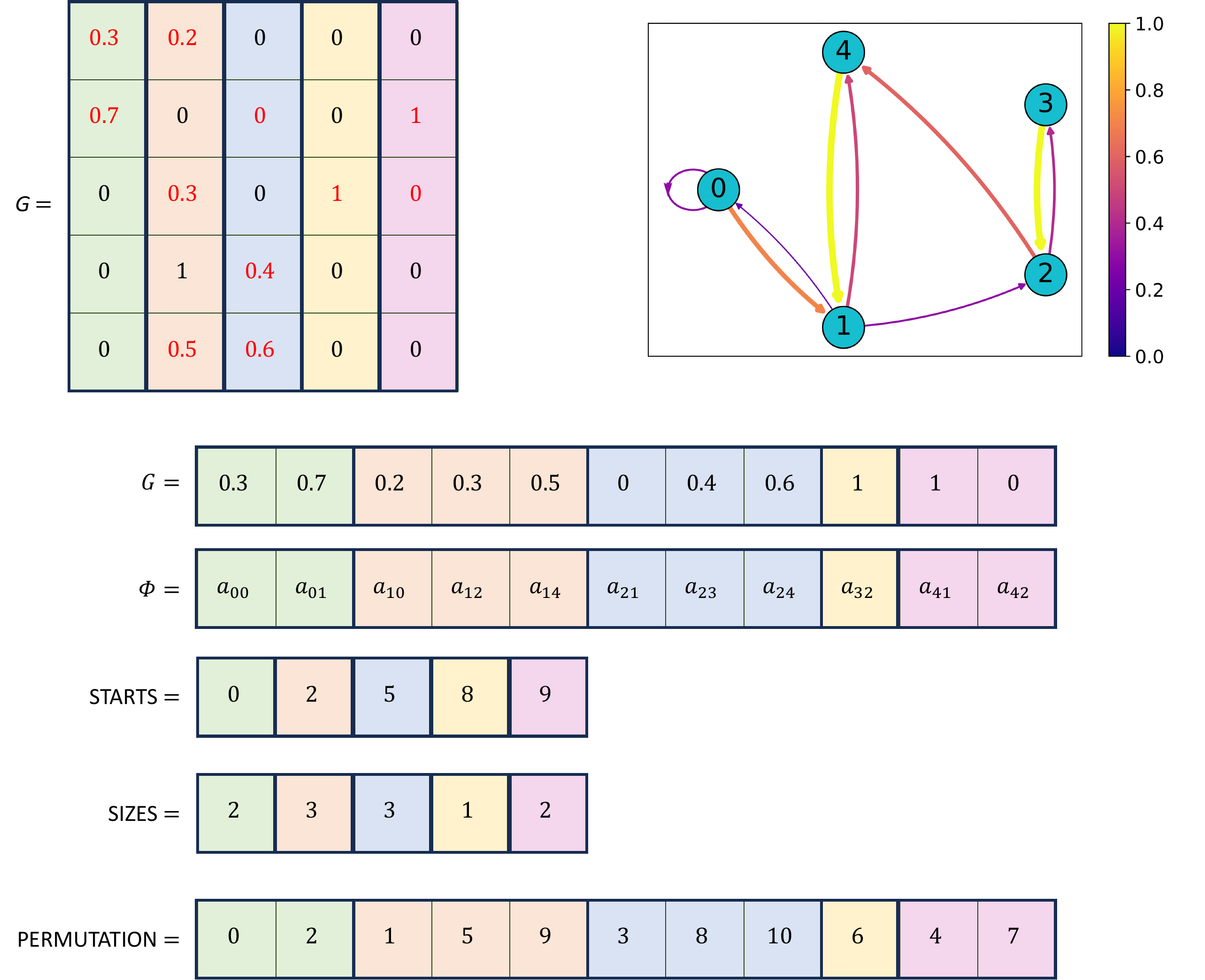}
    }
	\caption{Sparse representation of the reduced subspace $\mathcal{H}_R$ for a graph with $N=5$ nodes. Although in the dense representation the transition matrix has $N^2 = 25$ elements, the graph has only $9$ edges with non-null weights. Given the symmetry of the reduced subspace, it consists of $11$ elements ($5$ edges between different nodes and $1$ self-loop), whose associated positions in the transition matrix are shown in red. Each column of the transition matrix $\TM$ corresponds to the probabilities of edges outgoing each node, and concatenating the reduced elements we obtain the reduced vector representation, which is a vector with $11$ elements separated in $5$ blocks. The reduced vector $\Phi$ represents the elements of an arbitrary quantum state $\left|\phi\right>$ (19), in the reduced subspace. To simulate the quantum walk evolution using this representation, we need the STARTS list containing the initial position of each block, the SIZES list containing the size of each block, and the PERMUTATION list relating each element to its swapped edge.}
	\label{F:SM_space}
\end{figure}

In this section, we show how to represent the elements of the reduced subspace $\mathcal{H}_R$ in a sparse manner using NumPy. An example for a graph with $N=5$ nodes is shown in Figure \ref{F:SM_space}.

To take the computational basis elements that correspond to the reduced subspace, we use the transition matrix $\TM$. However, since the reduced subspace is formed according to the underlying undirected graph, we must actually look at the non-null elements of the transition matrix $\TM$ in a symmetric form. This is so because even if a state has null amplitude in the positions corresponding to null elements in the transition matrix, due to the swap operator $S$, after the evolution a non-null amplitude can appear on such positions. Since the transition matrix is column-stochastic by construction, each column corresponds to a computational basis state of the first register. Thus, we can divide the quantum states in $N$ blocks, corresponding to the selected positions in $\TM$. Concatenating these blocks into a single vector, which we denote as reduced vector, we actually obtain the representation of the state in the reduced subspace $\mathcal{H}_R$, thus reducing the memory requirements considerably for sparse graphs.

Although we have stored efficiently the relevant elements of the objects, we still need to relate them to the indexes in the original dense objects. However, since now all the mathematical objects involved in the evolution have the same reduced vector structure, they are aligned and we do not need to store different index information for each object as in a usual sparse library. In contrast, we just need to store the relation with the dense objects once in a global form. To perform all the simulation algorithms, we actually only need the following three objects:

\begin{itemize}
	\item STARTS: A list of $N$ elements indicating the initial position of each block.
	
	\item SIZES: A list of $N$ elements indicating the size of each block.
	
	\item PERMUTATION: A list of $N_R$ elements indicating for each element which one corresponds to its swapped edge.
\end{itemize}

\subsection{Expansion and Recomposition: Reflection and Update Operators}

In the main text we have seen the expression of the reflection operator $R$:
\begin{equation}\label{SM_reflection}
	R = 2\sum_{i=0}^{N-1} \left|\psi_i\right>\left<\psi_i\right| - \mathbbm{1}, \ \ \ \left|\psi_i\right> := \left|i\right>_1 \otimes \sum_{k=0}^{N-1} \sqrt{\TM_{ki}}\left|k\right>.
\end{equation}
and also provided two different constructions for the update operator $\update$:
\begin{equation}
	\update_1 = \sum_{i=0}^{N-1} \left|\psi_i\right>\left<i,0\right| + \sum_{i=0}^{N-1} \left|\psi_i^\perp\right>\left<i,0^\perp\right| + \identity - \sum_{i=0}^{N-1} \left|i,0\right>\left<i,0\right| - \sum_{i=0}^{N-1} \left|i,0^\perp\right>\left<i,0^\perp\right|.
\end{equation}
\begin{equation}
	\update_2 = \identity - 2\sum_{i=0}^{N-1}\left|i,v_i\right>\left<i,v_i\right|.
\end{equation}
All these operators have terms with a similar structure, which can be expressed in a generic form as
\begin{equation}\label{expansion_recomposition}
	\Pi_{XY} = \sum_{i=0}^{N-1} \left|y_i\right>\left<x_i\right|,
\end{equation}
where the states $\left|x_i\right>$ and $\left|y_i\right>$ are expressed as
\begin{equation}
	\left|x_i\right> := \left|i\right>_1 \otimes \sum_{k=0}^{N-1} X_{ki}\left|k\right>_2, \ \ \ \left|y_i\right> := \left|i\right>_1 \otimes \sum_{k=0}^{N-1} Y_{ki}\left|k\right>_2.
\end{equation}
Thus, they can be represented by two $N\times N$ matrices $X$ and $Y$. Moreover, these matrices share the same sparsity that the transition matrix $G$, and thus can also be represented as a reduced vector relating its elements to the computational basis states of the reduced subspace $\mathcal{H}_R$. For example, in the case of the reflection $R$, the matrix representing the states $\left|\psi_i\right>$ is obtained taking the element-wise square root of the transition matrix $G$.

\begin{figure}[h]
	\centering
	\makebox[10pt][c]{
	\includegraphics[scale=0.4]{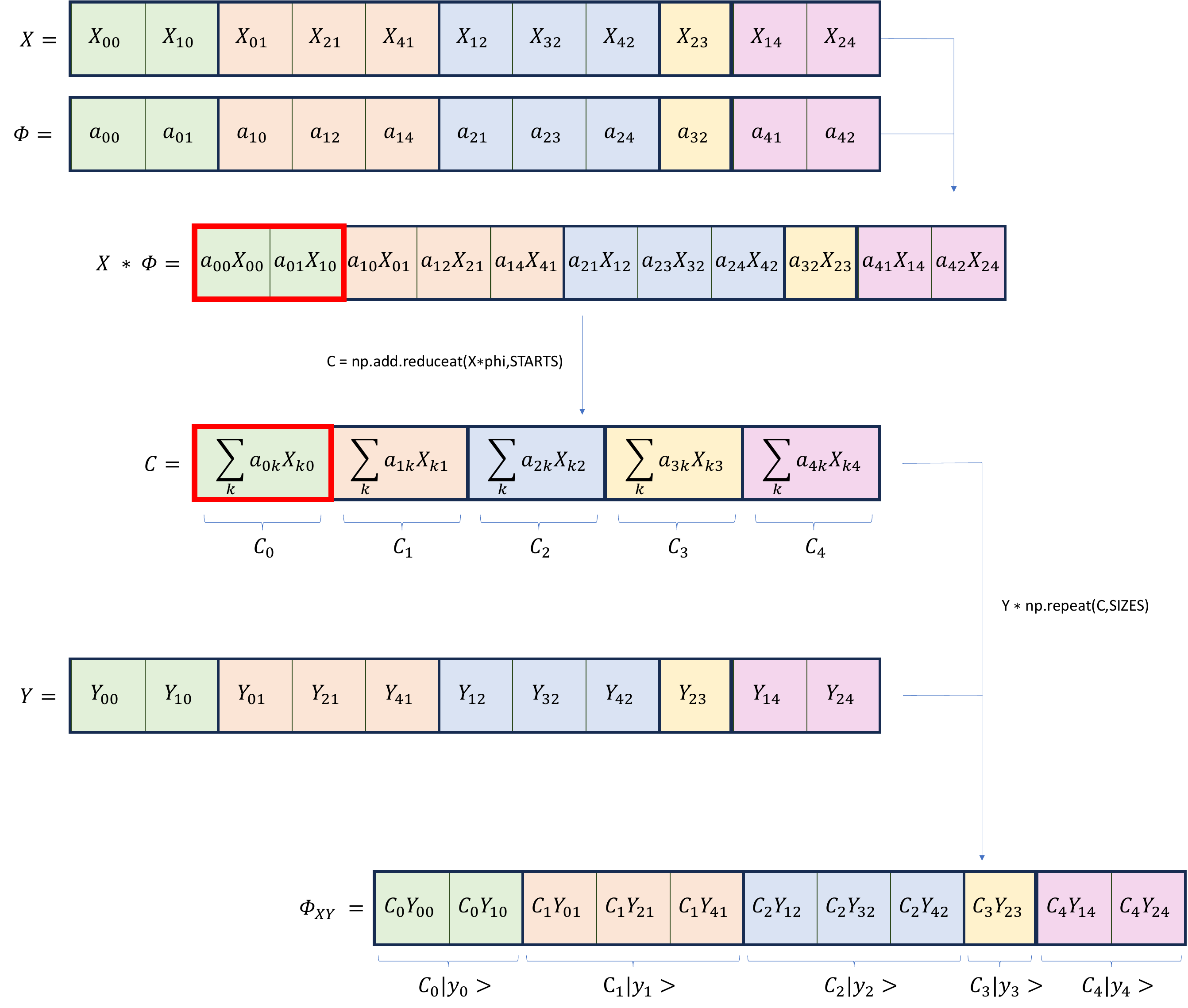}
    }
	\caption{Expansion-recomposition in the reduced subspace for a graph with $N=5$ nodes. For the expansion in the set of states $\left|x_i\right>$, the coefficients are calculated by the element-wise multiplication of the reduced vector $X$ and the reduced vector state $\Phi$, and then adding the elements inside each block. For the recomposition on a set of states $\left|y_i\right>$, the coefficients of the vector $C$ are repeated to match the reduced vector structure, and then multiplied element-wise with the reduced vector $Y$.}
	\label{F:SM_expansion-recomposition}
\end{figure}

To simulate the action of these operators, we need to simulate the action of the general term in \eqref{expansion_recomposition}, which corresponds to the primitives expansion and recomposition explained in Section III A 1 of the main text:
\begin{itemize}
	\item Expansion: It consists of obtaining the coefficients of a state $\left|\phi\right>$ along a set of states $\left|x_i\right>$:
	\begin{equation}
		C_i = \left<x_i\right.\left|\phi\right>.
	\end{equation}
	\item Recomposition: It consists of using some coefficients for reconstructing a state given a set of states $\left|y_i\right>$:
	\begin{equation}
		\left|\phi\right>_{XY} = \Pi_{XY}\left|\phi\right> = \sum_{i=0}^{N-1} C_i\left|y_i\right>.
	\end{equation}
\end{itemize}
In the main paper we described a method for simulating these two primitives taking advantage of the matrix representation of the state, so that we only perform element-wise multiplications, and also avoid explicit \textit{for} loops. In this case, we want to proceed in an analogue way, but taking into account the reduced vector representation. An example for a graph with $N=5$ nodes is shown in Figure \ref{F:SM_expansion-recomposition}.

For the expansion, we need to calculate all the dot products at the same time. To do so, we take the element-wise multiplication between the reduced vectors representing the state $\left|\phi\right>$ and the matrix $X$. After that, we need to add the elements of each block. To avoid \textit{for} loops, we use the \textit{add.reduceat} feature of NumPy, which allows to add in parallel each block using the initial positions of each of them contained in the list STARTS.

For the recomposition, we use the \textit{repeat} feature of NumPy, which provided the SIZES list, creates a vector repeating the coefficients to match the structure of blocks of the reduced vector. Thus, it can be multiplied element-wise with the reduced vector representing the matrix $Y$, recomposing the state $\left|\phi\right>_{XY}$.

With the expansion and recomposition primitives we can simulate the reflection $R$ and update $\update$ operators. However, note that in the case of the update operator, it acts in all the states of the form $\left|i,0\right>$. Thus, if we are to simulate this operator, in the case that these states are not present in the reduced subspace $\mathcal{H}_R$, we need to add them when we define it. In the worst case this means that we are adding $2N$ elements to each object. However this is an assumable cost that does not break the sparsity of the problem.

\newpage

\subsection{Swap}

The swap operator is defined as:
\begin{equation}\label{SM_swap}
	S := \sum_{i,j=0}^{N-1} \left|i,j\right>\left<j,i\right|.
\end{equation}

The action of this operator is a permutation of the reduced vector elements, so that the amplitude of each directed edge is swapped with the one of the inverse edge. Thus, to simulate it we just access to the elements of the reduced vector with the PERMUTATION list to reorganize it.

\subsection{Reflection Around $0$}

The reflection $R_0$ is defined as:
\begin{equation}
	R_0 = 2\sum_{i=0}^{N-1} \left|i,0\right>\left<i,0\right| - \identity.
\end{equation}
Its action corresponds to changing the sign of all the elements of the state vector, except those with state $\left|0\right>_2$ in the second register. This particular states in the reduced vector correspond to the initial element of each block. Thus, to simulate it, we just have to multiply the entire vector by $-1$, and then multiply by $-1$ the initial element of each block using the STARTS list.

Again, note that for implementing this operator we need that all the states $\left|i,0\right>$ exist in the reduced subspace, so that we have to add these edges the same as for the update operator $\update$.

\subsection{Oracle}

The unitary evolution operator of the quantum walk can be modified introducing oracles, for example in the context of quantum search. Let us define the operator $Q$ as follows:
\begin{equation}\label{oracle}
	Q = \mathbbm{1}_N - 2\sum_{k \in \mathcal{M}} \left|k\right>\left<k\right|,
\end{equation}
which inverts the sign of the vectors $\left|k\right>$ given a set $\mathcal{M}$ of nodes to mark. In order to mark the nodes in the first (second) register, we use the oracle $Q_1$ ($Q_2$) defined as:
\begin{equation}
	Q_1 = Q \otimes \mathbbm{1}_N, \ \ \ Q_2 = \mathbbm{1}_N \otimes Q.
\end{equation}
The action of the oracle $Q_1$ is a multiplication by $-1$ of all the elements whose computational basis state has a marked node in the first register. Thus, it actually corresponds to multiplying by $-1$ all the blocks corresponding to the marked nodes. Using the lists STARTS and SIZES we can create a list with the indexes whose sign must be flipped, and use it to simulate the action of the oracle. For the oracle on the second register, we just use the relationship $Q_2 = S Q_1 S$.

\subsection{Measurement}

The probability distribution of the walker being at each node after measuring the first register is obtained by adding the probability associated to each edge state for the $N$ blocks. Thus, we just have to take the element-wise squared modulus of the reduced vector, and use the STARTS list to add the elements of each block. Regarding a measurement in the second register, the probability distribution can be obtained in the same manner, as long as we first swap the elements using the PERMUTATION list.

\subsection{Vectorization on Several States}

If we stack different reduced vectors along a new axis, forming a tensor object, it turns out that all the NumPy-based methods that we have described above can be applied in a vectorized manner, so that the operations are performed in parallel for all the vectors. Thus, we can parallelize the evolution on different quantum states, the same as the dense simulation algorithms of the main paper.

\end{document}